\documentstyle[11pt]{article}

\def\dim{{\rm dim}}
\def\det{{\rm det}}\def\tr{{\rm Tr}}

\def\cF{{\cal F}}\def\cG{{\cal G}}\def\cH{{\cal H}}
\def\cL{{\cal L}}\def\cM{{\cal M}}\def\cN{{\cal N}}
\def\cP{{\cal P}}\def\cQ{{\cal Q}}
\def\cS{{\cal S}}\def\cZ{{\cal Z}}
\def\quat{{\bf H}}
\def\eqdef{\stackrel{\rm def}{=}}
\def\twomat#1#2#3#4{\left(\matrix{#1&#2\cr #3&#4\cr}\right)}

\def\twovec#1#2{\left(\matrix{#1\cr #2\cr}\right)}

\def\o#1#2{{#1\over#2}}

\def\um{{1 \over 2}}
\def\un{{\bf 1}}
\def\bx{{\bf X}}

\def\r{{\bf R}}
\def\c{{\bf C}}

\def\csg{\c^2/\Gamma}

\def\hika{HyperK\"ahler~}
\def\ka{K\"ahler~}

\def\cJ#1{{\cal J}^{#1}}
\def\cJt#1#2#3{{\cal J}^{#1}_{#2#3}}
\def\epsi#1#2#3{\varepsilon^{#1#2#3}}

\def\dim{{\rm dim}}
\def\det{{\rm det}}
\def\tr{{\rm Tr}}

\def\cF{{\cal F}}\def\cG{{\cal G}}\def\cH{{\cal H}}
\def\cL{{\cal L}}\def\cM{{\cal M}}\def\cN{{\cal N}}
\def\cP{{\cal P}}\def\cQ{{\cal Q}}
\def\cS{{\cal S}}\def\cZ{{\cal Z}}
\def\quat{{\bf H}}
\def\eqdef{\stackrel{\rm def}{=}}
\def\twomat#1#2#3#4{\left(\matrix{#1&#2\cr #3&#4\cr}\right)}

\def\twovec#1#2{\left(\matrix{#1\cr #2\cr}\right)}

\def\o#1#2{{#1\over#2}}

\def\um{{1 \over 2}}
\def\un{{\bf 1}}
\def\bx{{\bf X}}

\def\r{{\bf R}}
\def\c{{\bf C}}

\def\csg{\c^2/\Gamma}
\def\hika{HyperK\"ahler~}
\def\ka{K\"ahler~}

\def\f#1#2#3{f^{#1 #2 #3}}
\def\twomat#1#2#3#4{\left(\matrix{#1&#2\cr #3&#4\cr}\right)}

\def\twovec#1#2{\left(\matrix{#1\cr #2\cr}\right)}

\def\un{{\bf 1}}
\def\det{{\rm det}}\def\tr{{\rm Tr}}
\def\dag#1{{#1}^{\dagger}}
\def\dags#1#2{{#1}^{#2\dagger}}

\def\ps#1{\psi^{#1}}
\def\pss#1{\psi^{{#1}^*}}
\def\pst#1{{\tilde\psi}^{#1}}
\def\psts#1{{\tilde\psi}^{{#1}^*}}
\def\lap{\lambda^{\scriptscriptstyle +}}
\def\latp{{\tilde\lambda}^{\scriptscriptstyle +}}
\def\lamm{\tilde\lambda^{\scriptscriptstyle -}}
\def\latm{{\tilde\lambda}^{\scriptscriptstyle -}}
\def\mup{\mu^{\scriptscriptstyle +}}
\def\mutp{{\tilde\mu}^{\scriptscriptstyle +}}
\def\mum{\mu^{\scriptscriptstyle -}}
\def\mutm{{\tilde\mu}^{\scriptscriptstyle -}}
\def\latp{\lambda^{\scriptscriptstyle +}}
\def\lap{{\tilde\lambda}^{\scriptscriptstyle +}}
\def\latm{\lambda^{\scriptscriptstyle -}}

\def\mutp{\mu^{\scriptscriptstyle +}}
\def\mup{{\tilde\mu}^{\scriptscriptstyle +}}

\def\mum{\mu^{\scriptscriptstyle -}}
\def\mutm{{\tilde\mu}^{\scriptscriptstyle -}}
\def\psu#1{\psi_{\scriptscriptstyle u}^{#1}}
\def\psus#1{\psi_{\scriptscriptstyle u}^{{#1}^*}}
\def\psut#1{{\tilde\psi}_{\scriptscriptstyle u}^{#1}}
\def\psuts#1{{\tilde\psi}_{\scriptscriptstyle u}^{{#1}^*}}
\def\psv#1{\psi_{\scriptscriptstyle v}^{#1}}
\def\psvs#1{\psi_{\scriptscriptstyle v}^{{#1}^*}}
\def\psvt#1{{\tilde\psi}_{\scriptscriptstyle v}^{#1}}
\def\psvts#1{{\tilde\psi}_{\scriptscriptstyle v}^{{#1}^*}}

\def\delp{\nabla_{\scriptscriptstyle +}}
\def\delm{\nabla_{\scriptscriptstyle -}}
\def\dep{\partial_{\scriptscriptstyle +}}
\def\dem{\partial_{\scriptscriptstyle -}}
\def\cA{{\cal A}}\def\cD{{\cal D}}
\def\cF{{\cal F}}\def\cL{{\cal L}}\def\cP{{\cal P}}\def\cQ{{\cal Q}}
\def\cW{{\cal W}}
\def\cM{{\cal M}}\def\cN{{\cal N}}
\def\o#1#2{{#1\over #2}}
\def\eqdef{\stackrel{\rm def}{=}}
\def\twomat#1#2#3#4{\left(\matrix{#1&#2\cr #3&#4\cr}\right)}
\def\coco{{\rm \hskip 2pt c.c.\hskip 2pt}}
\def\ssp{{\scriptscriptstyle +}}
\def\ssm{{\scriptscriptstyle -}}

\def\o#1#2{{#1 \over #2}}

\begin{document}
\begin{flushright}
SISSA 187/94/EP\\
hep-th@xxx/9411183
\end{flushright}
\vskip 0.3cm
\begin{center}
{\Large\bf
HYPERKAHLER QUOTIENTS AND N=4 GAUGE THEORIES IN D=2
\footnote{\sl
Talk given by P. Fr\'e at the
{\it F. Gursey Memorial Conference}, Instanbul, June 1994}
 }
\end{center}
\vskip 0.2cm
\begin{center}
 Marco Bill\'o and  Pietro Fr\'e
\end{center}
\begin{center}
{\it SISSA - International School for Advanced Studies, via Beirut 2-4,
34140 Trieste, Italy \\ and I.N.F.N. - Sezione di Trieste - Trieste, Italy}
\end{center}
\vskip 2cm
\section{Introduction}
In this contribution we review some results we have recently
obtained on the relationship between certain geometrical constructions
of \hika manifolds and N=4 supersymmetry in D=2 space-time.
\hika manifolds are of interest in connection with
several issues in contemporary
theorethical physics:
\par {\it (i)} non-compact \hika manifolds in 4 dimensions
can be interpreted as gravitational instantons;
\par {\it (ii)}
4-dimensional $\sigma$-models with compact or
non-compact \hika $4n$-dimensional  target manifolds describe the interaction
of hypermultiplets in the N=2, D=4 supersymetric theories.
\par In particular
very recently the  role {\it (ii)} played by \hika manifolds has come
under attention
in relation with monopole theory, non-perturbative calculations in N=2
gauge theories and topological field theories in D=4
\cite{topf4d_1,topf4d_2,topf4d_3,topf4d_4,topf4d_5}.
\par From the point of view {\it (i)} an important question
not yet fully answered
concerns gravitational instanton effects in string theory.
Some progress has been made in constructing (4,4),
$c$=6 superconformal models
describing string propagation on such backgrounds
\cite{gravinstant_1,gravinstant_2};
yet one has still to go a long way in order to master this kind of problems.
Since string propagation on any manifold $\cM$ is described by a $2D$
$\sigma$-model with $\cM$ as target space it is clear that any
$2D$ field theory whose effective theory is a $\sigma$-model on a \hika $\cM$
has a deep meaning in the above context.
\par
Here we shall discuss precisely this kind of theories.
They are N=4 gauge theories
coupled to hypermultiplets with canonical kinetic terms, the only interaction
being induced by the minimal gauge coupling. Yet the structure of N=4
supersymmetry provides the link with the geometrical set-up of \hika quotients.
The basic result \cite{topphases_1} is that the auxiliary fields
of the gauge multiplet realize
in a lagrangian way the triholomorphic momentum map and lead
to an effective theory
that has the \hika quotient as target space.
What we have found is the N=4 generalization of a pattern discovered by Witten
in N=2,D=2 gauge theories. The essential difference is that while the N=2
case displays a two-phase structure, a $\sigma$-model phase  and a
Landau-Ginzburg phase, the N=4 case admits only the $\sigma$-model phase.
\par
In this framework and using the geometrical results of Kronheimer
\cite{momentmap_5,gravinstant_3} (reviewed in this note)
we show how to construct
a {\it microscopic field theory} that admits as
effective theory the N=4 $\sigma$-model on all
Asymptotically Locally Euclidean (ALE) self-dual spaces.
The physical data contained in the microscopic lagrangian
(the Fayet-Iliopoulos parameters \cite{miscellaneous_1})
are put by our construction into {\it explicit} correspondence
with the geometrical data associated with the ALE manifold, in particular the
moduli of the complex structures. This correspondence is believed to be of
relevance in  the topological theories obtained by twisting either
our microscopic
two-dimensional theory \cite{topphases_1}
 or N=2 supergravity in four dimensions
\cite{topftwist_1,topftwist_2,topf4d_1}.
\section{HyperkK\"ahler quotients}
\hskip 20pt
\underline{\sl \hika manifolds}
\par\noindent
On a \hika manifold $\cM$, which is necessarily $4n$-dimensional,
there exist three covariantly
constant complex structures $\cJ i: T\cM\rightarrow T\cM$, $i=1,2,3$;
the metric is hermitean with respect to all of them and they satisfy the
quaternionic algebra: $\cJ i \cJ j = - \delta^{ij} + \epsi ijk \cJ k$.
\par
In a vierbein basis $\{V^A\}$, hermiticity of the metric is equivalent
to the statement that  the matrices $\cJt iAB$ are
antisymmetric. By covariant constancy, the three \hika
two-forms $\Omega^i=\cJt iAB V^A\wedge V^B$ are closed: $d\Omega^i=0$.
In the four-dimensional case, because of the quaternionic algebra
constraint, the $\cJt iAB$
can  be either selfdual or
antiselfdual; if we take them to be antiselfdual: $\cJt iAB =-{1\over 2}
\epsilon_{ABCD} \cJt iCD$, then the integrability condition for the
covariant constancy of $\cJ i$ forces the curvature two-form $R^{AB}$ to
be selfdual: thus, in the four-dimensional case, \hika manifolds are
particular instances of gravitational instantons.
\par
A \hika manifold is a \ka manifold with respect to each of
its complex structures.
\par
\underline{\sl Momentum map}
\par\noindent
Consider a compact Lie group $G$ acting on a \hika manifold $\cS$ of
real dimension $4n$ by means of Killing vector fields $\bx$ that are
holomorphic
with respect to the three complex structures of $\cS$; then these vector
fields preserve also the \hika forms:
\begin{equation}
\left.\begin{array}{l}
{\cal L}_{\scriptscriptstyle\bx}g = 0 \leftrightarrow
\nabla_{(\mu}X_{\nu)}=0 \\
{\cal L}_{\scriptscriptstyle\bx}\cJ i = 0 \,\, , \,i =1,2,3\\
\end{array}\right\} \,\,\Rightarrow\,\, 0={\cal L}_{\scriptscriptstyle\bx}
\Omega^i = i_{\scriptscriptstyle\bx} d\Omega^i+d(i_{\scriptscriptstyle\bx}
\Omega^i) = d(i_{\scriptscriptstyle\bx}\Omega^i)\, .
\label{holkillingvectors}
\end{equation}
Here ${\cal L}_{\scriptscriptstyle\bx}$ and
$i_{\scriptscriptstyle\bx}$ denote respectively the Lie derivative along
the vector field $\bx$ and the contraction (of forms) with it.

If $\cS$ is simply connected, $d(i_{\bx}\Omega^i)=0$ implies the existence
of three functions ${\cal D}_i^{\bx}$ such that $d{\cal D}_i^{\bx}=
i_{\scriptscriptstyle\bx}\Omega^i$. The functions ${\cal D}_i^{\bx}$ are
defined up to a constant,
which can be arranged so to make them equivariant: $\bx {\cal D}_i^{\bf Y} =
{\cal D}_i^{[\bx,{\bf Y}]}$.

The $\{{\cal D}_i^{\bx}\}$ constitute then a {\it momentum
map}. This can be regarded as a map ${\cal D}: \cS \rightarrow \r^3\otimes
\cG^*$, where $\cG^*$ denotes the dual of the Lie algebra $\cG$ of the
group $G$.
Indeed let $x\in \cG$ be the Lie algebra element corresponding to the Killing
vector $\bx$; then, for a given $m\in \cS$, ${\cal D}_i(m) :x\longmapsto
{\cal D}_i^{\bx}(m)\in\c$ is a linear functional on $\cG$. In practice,
expanding $\bx =X_a {\bf k}^a$ in a basis of Killing vectors ${\bf k}^{a}$
such that $[{\bf k}^a,{\bf k}^b]=\f abc {\bf k}^c$, where
$\f abc $ are the structure constants of $\cG$,
we also have  ${\cal D}_i^{\bx}=X_a \, {\cal D}_i^a$, $i=1,2,3$;
the $\{{\cal D}_i^a\}$ are the components of the momentum map.
\par
\underline{\sl \hika quotient}\par\noindent
It is a procedure that provides a way to construct from
$\cS$ a
lower-dimensional \hika manifold $\cM$, as follows.
Let ${\cZ}^*\subset \cG^*$ be
the dual of the centre of $\cG$. For each $\zeta\in \r^3\otimes \cZ^*$ the
level set of the momentum map
\begin{equation}
\cN \equiv\bigcap_{i} {\cal D}_i^{-1}(\zeta^i) \subset \cS ,
\label{levelsetdef}
\end{equation}
which has dimension $\,\dim\,\cN=\dim\,\cS -3\,\,\dim\,G$,
is invariant under the action of $G$, due to the equivariance of
${\cal D}$. It is thus possible to take the quotient
\[\cM =\cN/G.\]
$\cM$ is a smooth manifold of dimension ${\rm dim}\cM={\rm dim}\cS
-4\, {\rm dim}G$ as long as the action of $G$ on $\cN$ has no fixed points.
The three two-forms $\omega^i$ on $\cM$, defined  via
the restriction to $\cN\subset\cS$ of the $\Omega^i$ and the quotient
projection from $\cN$ to $\cM$,
are closed and satisfy the quaternionic algebra thus
providing $\cM$ with a \hika structure.
\par
Once $\cJ 3$ is chosen as
the preferred complex structure, the momentum maps
${\cal D}_{\pm}={\cal D}_1\pm i{\cal D}_2$
are holomorphic (resp. antiholomorphic) functions.
\par
The standard use of the \hika quotient is that of obtaining non trivial
\hika manifolds starting from a
flat $4n$ real-dimensional manifold $\r^{4n}$ acted on by a suitable
group G generating triholomorphic isometries
\cite{momentmap_2,hyperquotient_2}.
\par
This is the way it was utilized by Kronheimer
\cite{momentmap_5,hyperquotient_3}
in its exhaustive
construction of all self-dual asymptotically locally Euclidean four-spaces
(ALE manifolds) \cite{momentmap_4,gravinstant_3,gravinstant_4,gravinstant_5},
that we consider later.

Indeed the manifold $\r^{4n}$ can be given a quaternionic structure, and the
corresponding quaternionic notation is sometimes convenient. For $n=1$
one has the flat quaternionic space \label{quater}
$\quat\stackrel{\rm def}{=}\left(\r^4,\left\{J^i\right\}\right)$ . We represent
its elements
\[ q\in\quat=x+i y+j z+k t=x^0+x^i J^i,\hskip 15pt x,y,z,t\in\r\]
realizing the quaternionic structures $J^i$ by means of Pauli matrices:
 $J^i=i\left(\sigma^i\right)^T$. Thus
\begin{equation}
q=\twomat{u}{iv^*}{iv}{u^*} \hskip 1cm\longrightarrow\hskip 1cm
\dag q=\twomat{u^*}{-iv^*}{-iv}{u}
\label{singlequaternion}
\end{equation}
where $u=x^0+ix^3$ and $v=x^1+ix^2$.
The euclidean metric on $\r^4$ is retrieved as $d\dag q\otimes dq=ds^2
\un$. The \hika forms are grouped into a quaternionic two-form
\begin{equation}
\Theta=d\dag q\wedge dq\,\,\stackrel{def}{=}\,\,\Omega^i J^i=
\twomat{i\Omega^3}{i\Omega^+}{i\Omega^-}{-i\Omega^3}\,\, .
\label{thetaform}
\end{equation}
For generic $n$, we have the space $\quat^n$, of elements
\begin{equation}
q=\twomat{u^A}{iv^{A^*}}{i v^A}{u^{A^*}} \hskip 1cm
\longrightarrow\hskip 1cm \dag q=\twomat{u^{A^*}}{-iv^{A^*}}
{-iv^A}{u^A}\hskip 1cm
\begin{array}{l}u^A,v^A\in\c^n\\A=1,\ldots n\end{array}
\label{multiquaternion1}
\end{equation}
Thus $d\dag q\otimes dq=ds^2 \un$ gives $ds^2=d
u^{A^*}\otimes du^A+dv^{A^*}\otimes dv^A$ and the \hika forms are grouped
into the obvious generalization of the
quaternionic two-form in eq.(\ref{thetaform}):
$\Theta=\sum_{A=1}^n d\dags qA \wedge dq^A=\Omega^i J^i$,
leading to $\Omega^3=2i\partial\bar\partial K$
where the \ka potential $K$ is $K=\um\left(u^{A^*} u^A + v^{A^*}
v^A\right)$,
and to $\Omega^+=2 i du^A\wedge dv^A$,
$\Omega^-=\left(\Omega^+\right)^*$.
\par
Let $\left(T_a\right)^A_B$ be the antihermitean generators of a compact
Lie group G in its  $n\times n$ representation. A triholomorphic action
of $G$ on $\quat^n$ is realized by the Killing vectors of components
\begin{equation}
X_a=\left(\hat T_a\right)^A_B q^B {\partial\over\partial q^A}+\dags qB
\left(\hat T_a\right)^B_A {\partial\over\partial \dags qA}\hskip 1cm ;
\hskip 1cm \left(\hat T_a\right)^A_B=\twomat{\left(T_a\right)^A_B}{{\bf
0}}{{\bf 0}}{\left(T^*_a\right)^A_B}\,\, .
\label{triholoaction}
\end{equation}
Indeed one has $\cL_{\scriptscriptstyle \bx}\Theta=0$.
The corresponding components of the momentum map are:
\begin{equation}
{\cal D}^a=\dags qA\twomat{\left(T_a\right)^A_B}{{\bf
0}}{{\bf 0}}{\left(T^*_a\right)^A_B}q^B +\twomat{c}{\bar b}{b}{-i c}
\label{momentumcomponents}
\end{equation}
where $c\in\r,\,b\in\c$ are constants.
\section{\hika quotients in D=2, N=4 theories}
Consider a supersymmetric $\sigma$-model from a 2-dimensional N=4 super-world
sheet to a target space $\cS$. It is well known that supersymmetry requires
$\cS$ to be \hika. Introduce also the supersymmetric gauge multiplet for a
group
$G$ acting triholomorpiycally on $\cS$. It turns out that the auxiliary
fields of the gauge multiplet $\{\cP,\cQ\}$, $(\cP\in {\bf R},\cQ\in {\bf  C})$
are identified with the momentum functions $\{\cD^3,\cD^{\pm}\}$for the
$G$-action on $\cS$.
\par
In view of this fundamental property, the \hika quotient
offers a natural way to construct a  N=4, D=2 $\sigma$-model
on a non-trivial manifold $\cM$ starting from a free
$\sigma$-model on a flat-manifold ${\cal S}={\bf H}^n$.
It suffices to gauge appropriate triholomorphic
isometries by means of non-propagating gauge multiplets.
Omitting the kinetic term of these gauge multiplets
and performing the gaussian integration of
the corresponding fields one realizes the \hika quotient in a Lagrangian way.
\par
Actually the \hika quotient is a generalization of a similar
\ka quotient procedure, where the momentum map
${\cal D}: \cS \rightarrow \r\otimes
\cG^*$ consists  just of one hamiltonian function, rather than three.
The \ka quotient is
related with N=2,D=2 supersymmetry, the reason being that in this case
the vector multiplet contains just one real auxiliary field  $\cP$.
\par
Recently, Witten has reconsidered the
\ka quotient construction of an N=2 two-dimensional
$\sigma$-model in \cite{topphases_2}.
We constructed in \cite{topphases_1} the analogue N=4  model, and compared it
with the N=2 model.
We review here some fundamental points of this work.
\par
\underline{\sl Results for the N=2 theory}
\par\noindent
In the  N=2 case a vector multiplet is composed of a gauge boson ${\cal A}$,
namely a world-sheet 1-form,
two spin 1/2 gauginos, whose four components
we denote by $\lambda^+$,$\lambda^-$,
$\tilde\lambda^+$,$\tilde\lambda^-$,
a complex physical scalar $M $ and a real
auxiliary scalar ${\cal P}$.
For simplicity we write here the formul\ae~ for a U(1) gauge group
\footnote{The extension to several abelian groups
is trivial and for a generic group
it is contained in \cite{topphases_1} in the same setting as here.}
and we consider a linear superpotential \cite{topphases_2}. $ i {t\over 4} M$,
with the coupling $t$ is $t = i r + {\theta \over 2\pi} $.
Denote $\partial_{\pm} = \partial /\partial z^o \pm
\partial /\partial z^1$, $z^{\alpha}$ being the
world-sheet coordinates.
Then we have:
\begin{eqnarray}
\cL^{(2)}_{\rm gauge} & = &\o 12 \cF^2 -i\,\bigl(\lap\dep\lamm +\latp\dem\latm
\bigr) -4\bigl(\dep M^*\dem M +\dem M^*\dep M)\nonumber\\
&&\mbox{}+ 2\cP^2 -2r\cP +\o{\theta}{2\pi}\cF
\label{ntwo9}
\end{eqnarray}
$r\cP$ is the Fayet-Iliopoulos term \cite{miscellaneous_1},
$\theta\cF /2\pi$ a topological term.
We consider then $n$ chiral multiplets  $X^i,\ps i, \pst i,\cH_i$
(plus their complex conjugates) of charge $q^i$
with respect to the above $U(1)$ group.
The complex scalars $X^i$ span ${\bf C}^n$.
The complex auxiliary field $\cH^i$ is identified with the derivative
of a holomorphic superpotential $W(X)$: $\cH^i = \partial_{i^*} W^*$.
The supersymmetric lagrangian is
\begin{eqnarray}
\lefteqn{\cL^{(2)}_{\rm chiral}\, =\,
-\bigl(\delp X^{i^*}\delm X^i + \delm X^{i^*}\delp X^i
\bigr) + 4i\bigr(\ps i\delm\pss i +\pst i\delp\psts i\bigr) }\nonumber \\
&&\mbox{}+ 8\biggl(
\bigl(\ps i\pst j\partial_i\partial_j W +\coco\bigr) +
\partial_i W\partial_{i^*}W^*\biggr) +2i\,\sum_i q^i\bigl(
\ps i\lamm X^{i^*} -\pst i\latm X^{i^*} -\coco\bigr)\nonumber\\
&&\mbox{}+8i\,\biggl(M^*\sum_i q^i\ps i\psts i -\coco\biggr) +8
M^*M\,\sum_i (q^i)^2\,X^{i^*} X^i -2\cP\sum_i q^i X^{i^*}X^i
\label{ntwo14}
\end{eqnarray}
Here and henceforth $\nabla$ is the covariant derivative constructed by
means of the gauge connection $\cA$.\par
The total lagrangian is $\cL^{(2)}\,=
\,\cL^{(2)}_{\rm gauge}$ $ -\cL^{(2)}_{\rm chiral}$,
the relative sign being fixed by the requirement of positivity of the energy.
Note that ${\cal D}^{\bf X}\left ( X,X^*\right )
\,= \,\sum_i q^i |X^i|^2$ is the momentum map function
for the holomorphic action of the
gauge group on the matter multiplets.
Eliminating $\cP$ through its own equation of motion we get:
$\cP\, = \, -\o 12 (D^{\bf X}(X,X^*) -r)$
and the bosonic potential reduces to:
\begin{equation}
U\,=\,\o 12 \biggl[r - \sum_i q^i |X^i|^2\biggr]^2 +8|\partial_i
W|^2 +8 |M|^2\sum_i (q^i)^2|X^i|^2
\label{ntwo19}
\end{equation}
Let us  focus now on the low energy effective theory for this model.
First of all consider the
structure of the classical vacua.
We must extremize the potential (\ref{ntwo19}).
Witten considered \cite{topphases_2} the case with L.G. potential
$ W = X^0 \, {\cal W}(X^i)$ where $\cW(X^i)$
is quasi homogeneous of degree $d$ if the $X^i$'s
are assigned as homogeinity weigths their charges $q^i$, and is transverse:
$\partial_i{\cal W} =0 \hskip 3pt\forall i$
iff $X^i =0\hskip 3pt\forall i$. $X^0$ has charge $-d$. Then two phases appear:
\begin{itemize}
\item $r > 0$: {\sl $\sigma$-model phase}. The space of classical vacua
is a transverse hypersurface embedded in ${\bf WCP}^N_{q^1\ldots q^N}$:
$X^0 = M = 0$, $\sum_i q^i|X^i|^2 = r$,  ${\cal W}(X^i)=0$.
\item $r < 0$: {\sl Landau-Ginzburg phase}. The space of vacua is a point:
$X^0 = \sqrt{\o{-r}d}$, $X^i = M = 0$.
The low-energy theory describes massles fields governed
by a Landau-Ginzburg potential ${\cal W}(X^i)$
\end{itemize}
Our interest is in the low-energy theory around a
vacuum of the first type. This theory turns
out  to be correctly described by a N=2 $\sigma$-model
\cite{topphases_1}. It emerges via
the lagrangian realization of a K\"ahler quotient.
Here we look just at the bosonic fields.
To be simple and definite, consider the ${\bf CP}^n$ model,
that is the case $q^A=1$, $A = 0,1,\ldots,n$
and $W(X^A)=0$. In this case there is only the $\sigma$-model phase.
\par
 Reinstall the gauge coupling constant
$g$ in the lagrangian $\cL^{(2)}$. Then let $g \rightarrow \infty$.
We are left with a gauge invariant lagrangian describing
matter coupled to gauge fields that  have no kinetic terms. Then we vary the
action in these fields. The resulting equations of motion  express
the gauge fields in terms of the matter fields.
This procedure is nothing else, from the functional
integral viewpoint,  but the
gaussian integration over the gauge multiplet in the limit
$g \, \longrightarrow \, \infty$.
It amounts to
deriving the low-energy effective action around the
classical vacua of the complete,
gauge plus matter  system. Indeed we have seen that around these vacua the
oscillations of the gauge fields are massive, and thus decouple from the
low-energy point of view.
\par
In our example, the variation in the gauge connection
components  $\cA_{\pm}$ and in $M$ identifies them in terms of
the matter fields. In particular
$\cA_{\pm} = -{\rm i}(X\partial_{\pm} X^* -X^*\partial_{\pm} X)
/ 2 X^* X$.
This has to be substituted into the lagrangian.
This latter is by construction invariant under the
U(1) transformation $X^A \rightarrow e^{i\Phi} X^A$, $\Phi\in{\bf R}$.
Allow  now $\Phi\in{\bf C}$, thus complexifying
U(1) to ${\bf C}^* = {\bf C} -\{ 0 \}$. Introduce an extra field
$v$, transforming under ${\bf C}^*$ as
$v\rightarrow v+\o i2 (\Phi -\Phi^*)$. Then the combinations
$e^{-v} X^A$ undergo just a
$U(1)$ transformation: $e^{-v} X^A \rightarrow  e^{i {\rm Re}\Phi} e^{-v} X^A$.

By substituting in the lagrangian $X^A \rightarrow e^{-v} X^A$ it
becomes ${\bf C}^*$-invariant. In particular the term involving $\cP$
becomes $-2\cP (r-e^{-2v}X^*X)$. Performing the variation with respect  to
the auxiliary field $\cP$,  the resulting equation of motion
identifies the extra scalar field $v$ in terms of the matter fields.
Introducing $\rho^2 \equiv r$
the result is that $e^{-v}=\o {\rho}{\sqrt{X^*X}}$.

What is the geometrical meaning of the above ``tricks'' (introduction
of the extra field $v$, consideration of the complexified gauge group)?
The answer relies on the properties of the K\"ahler quotient construction
\cite{momentmap_2};
Let us recall a few concepts, using notions and
notations introduced in section 1
\par
Let ${\bf Y}(s) =Y^a{\bf k}_a (s)$ be a Killing vector on $\cS$ (in our case
${\bf C}^{N+1}$), belonging to $\cG$ (in our case ${\bf R}$), the algebra of
the gauge group. In our case ${\bf Y}$ has a single component: ${\bf Y}=i\Phi
(X^A \o{\partial \phantom{X^A}}{\partial X^A} -X^{A^*}\o{\partial
\phantom{X^{A^*}}}{\partial X^{A^*}})$ ($\Phi\in{\bf R}$). The $X^A$'s are the
coordinates on $\cS$.
Consider the vector field $I{\bf Y}\in\cG^c$ (the complexified algebra), $I$
being the complex structure acting on $T\cS$. In our case $I{\bf Y}=
\Phi(X^A \o{\partial\phantom{X^A}}{\partial X^A} + X^{A^*} \o{\partial
\phantom{X^{A^*}}}{\partial X^{A^*}} )$.
This vector field is orthogonal to the hypersurface
$\cD^{-1}(\zeta)$, for any level $\zeta$; that is, it generates transformations
that change the level of the surface. In our case the surface $\cD^{-1}
(\rho^2)\in {\bf C}^{N+1}$ is defined by the equation $X^{A^*}X^A =\rho^2$.
The infinitesimal transformation generated by $I{\bf Y}$ is $X^A\rightarrow
(1+\Phi)X^A$, $X^{A^*}\rightarrow (1+\Phi)X^{A^*}$ so that the transformed
$X^A$'s satisfy $X^{A^*}X^A = (1+2\Phi)\rho^2$.
As recalled in section I, the K\"ahler quotient consists in starting
from $\cS$, restricting to $\cN =\cD^{-1}(\zeta)$ and taking the quotient
$\cM =\cN /G$. The above remarks about the action of the complexified gauge
group suggest that this is equivalent (at least if we skip the problems
due to the non-compactness of $G^c$) to simply taking the quotient $\cS /
G^c$, the so-called ``algebro-geometric'' quotient
\cite{momentmap_2,momentmap_1}.
\par
The K\"ahler quotient allows in principle to determine the expression of the
K\"ahler form on $\cM$ in terms of the original one on $\cS$. Schematically,
let $j$ be the inclusion map of $\cN$ into $\cS$, $p$ the projection from
$\cN$ to the quotient $\cM=\cN /G$, $\Omega$ the K\"ahler form on $\cS$ and
$\omega$ the K\"ahler form on $\cM$. It can be shown \cite{momentmap_2} that
\begin{eqnarray}
\label{kq1}
\cS \hskip 3pt \stackrel{j}{\longleftarrow}\hskip 3pt & \cN =\cD^{-1}(\zeta) &
\hskip 3pt\stackrel{p}{\longrightarrow}\hskip 3pt \cM =\cN /G\nonumber\\
\Omega \hskip 3pt \longrightarrow \hskip 3pt & j^*\Omega =p^*\omega &
\hskip 3pt\longleftarrow\hskip 3pt \omega
\end{eqnarray}
In the algebro-geometric setting,
the holomorphic map that associates to a point
$s\in\cS$ (for us, $\{X^A\}\in{\bf C}^{N+1}$) its image $m\in\cM$ is obtained
as follows:\par\noindent
{\it i)} Bringing $s$ to $\cN$ by means of the finite action infinitesimally
generated by a vector field of the form ${\bf V} =I{\bf Y} = V^a{\bf k}_a$
\begin{equation}
\label{kq2}
\pi:\hskip 10pt s\in\cS\hskip 3pt\longrightarrow e^{-V} s\in\cD^{-1}(\zeta)
\end{equation}
{\it ii)} Projecting $e^{-V}$ to its image in the quotient $\cM =\cN /G$.\par
Thus we can consider  the pullback of the K\"ahler form $\omega$ through
the map $p\cdot\pi$:
\begin{eqnarray}
\label{kq3}
\cS\hskip 3pt\stackrel{\pi}{\longrightarrow}\hskip 3pt & \cN =\cD^{-1}(\zeta)
&\hskip 3pt\stackrel{p}{\longrightarrow} \hskip 3pt\cN /G\nonumber\\
\pi^* p^*\omega \hskip 3pt \longleftarrow \hskip 3pt & p^*\omega &
\hskip 3pt \longleftarrow\hskip 3pt \omega
\end{eqnarray}
Looking at (\ref{kq1}) we see that $\pi^* p^*\omega =\pi^* j^*\Omega$ so that
at the end of the day, in order to recover the pullback of $\omega$ to $\cS$
it is sufficient:
\par\noindent
{\it \phantom{i}i)} to restrict $\Omega$ to $\cN$\par\noindent
{\it ii)} to pull back this restriction to $\cM$ with respect to the map $\pi
=e^{-V}$.\par
We see from (\ref{kq2}) that the components of the vector field ${\bf V}$
must be determined by requiring
\begin{equation}
\label{kq2bis}
\cD(e^{-V}s)=\zeta
\end{equation}
In the lagrangian context,after having introduced the extra field
$v$ (which is now interpreted as the unique component of the vector field
${\bf V}$) to make the lagrangian ${\bf C}^*$-invariant,
eq. (\ref{kq2bis}) is retrieved as the equation of motion for $\cP$.
We have determined the form of the map $\pi$ : it
corresponds on the bosonic fields to $X^A\rightarrow e^{-v} X^A$.
\par
The remaining steps in
treating the lagrangian just consist in implementing
the K\"ahler quotient as in (\ref{kq3}). At the end
we obtain the $\sigma$-model on the target space $\cM$ (in our case
${\bf CP}^N$) endowed with the K\"ahler metric corresponding to the
K\"ahler form $\omega$. In our example such metric is the Fubini-Study
metric. Indeed in full generality one can show  \cite{momentmap_2}
that the  K\"ahler potential $\hat K$ for the manifold $\cM$, such that
$\omega= 2i\partial\bar\partial\hat K$ is given by
\begin{equation}
\label{kq4}
\hat K = K|_{\cN} + V^a \zeta_a
\end{equation}
Here $K$ is the K\"ahler potential on $\cS$; $K|_{\cN}$ is
its restriction to $\cN$, that is, it is computed after acting on the
point $s\in\cS$ with the transformation $e^{-V}$ determined by eq.
(\ref {kq2bis}); $V^a$ are the components of the vector field ${\bf V}$
along the $a^{\rm th}$ generator of the gauge group, and $\zeta_a$ those
of the level $\zeta$ of the momentum map; recall that the $\zeta$ belong to the
dual of the center, $\cZ^*$ and
therefore only the components of ${\bf V}$ along the
center actually contribute to eq. (\ref{kq4}). In our case we have the single
component $v$ given by $e^{-v} = \rho^2/\sqrt{X^* X}$; $\rho^2$ is the
single component of the level. The original K\"ahler potential on $\cS =
{\bf C}^{N+1}$ is $K = \o 12 X^{A^*}X^A$ so that when restricted to $\cD^{-1}
(\rho^2)$ it takes an irrelevant constant value $\o{\rho^2}2$. Thus
we deduce from (\ref{kq4}) that the K\"ahler potential for $\cM =
{\bf CP}^N$ that we obtain is $\hat K =\o 12\rho^2 \log (X^*X)$.
Fixing a particular gauge to perform the quotient with respect to
${\bf C}^*$ (see later),
this  potential can be rewritten as $\hat K =\o 12\rho^2\log (1+
x^*x)$, namely the Fubini-Study potential.

Indeed it is trivial to rewrite the lagrangian after
the substitution $X^A\rightarrow (\rho^2/\sqrt{X^* X}) X^A$.
We can then  utilize the gauge invariance to fix for instance,
in the coordinate patch where $X^0 \neq 0$, $X^0 = 1$.
That is, we fix completely
the gauge  going  from the homogeneous coordinates $(X^0,X^i)$ to the
inhomogeneous coordinates $(1, x^i=X^i/X^0)$ on ${\bf CP}^N$.\par
Having chosen our gauge, we rewrite the lagrangian
in terms of the fields $x^i$ .
Thus the bosonic lagrangian reduces finally
to that of a $\sigma$-model on ${\bf CP}^n$:
\begin{equation}
\cL = g_{i j^*} (\dep x^i\dem x^{j^*} + \dem x^i\dep x^{j^*})
\end{equation}
where $g_{i j^*}$ is the Fubiny-Study metric,
$g_{i j^*} = \o{\rho^2}{1+x^*x}\biggl(\delta_{ij} -\o{x^{i^*}x^j}{1+
   x^*x}\biggr)$.
\par
\underline{\sl The N=4 theory}
\par\noindent
The N=4 vector multiplet
\footnote{As for N=2 we write the formulae for a
U(1) gauge group and we refer for extension to a group $G$ to
\cite{topphases_1}.}, in addition to the gauge boson,
namely the 1-form ${\cal A}$,
contains four spin 1/2 gauginos whose eight
components are denoted by $\lambda^+$,$\lambda^-$,
$\tilde\lambda^+$,$\tilde\lambda^-$,$\mu^+$,$\mu^-$,
$\tilde\mu^+$,$\tilde\mu^-$, two complex physical scalars
$M\ne M^*$, $N\ne N^*$, and
three  auxiliary fields arranged into a real scalar
$\cP = {\cP}^*$ and a complex
scalar $\cQ \ne{\cQ}^*$.
\begin{eqnarray}
\cL^{(4)}_{\rm gauge} & = &\o 12 \cF^2 -i\bigl(\lap\,\dep\lamm +\mup\,\dep\mum
+\latp\,\dem\latm +\mutp\,\dem\mutm\bigr)\nonumber\\
&&\mbox{}+4\bigl(\dep M^*\,\dem M
+\dem M^*\,\dep M +\dep N^*\,\dem N +\dem N^*\,\dep N\bigr)\nonumber\\
&&\mbox{}+\o{\theta}{2\pi}\cF
+2\cP^2 +2\cQ^*\cQ -2r\cP-\bigl(s\cQ^* +s^*\cQ\bigr)
\label{nfour4}
\end{eqnarray}
Note that in addition to the Fayet-Iliopoulos and the topological term
present in the N=2 case we have term linear in the complex auxiliary
field $\cQ$, involving a new complex parameter $s$.

The quaternionic
hypermultiplets are the N=4 analogues of the N=2 chiral multiplets.
They are described by
a set of bosonic complex fields $u^i$, $v^i$, that can be
organized in quaternions
\begin{equation}
Y^{i}=\twomat{u^i}{iv^{i^*}}{iv^i}{u^{i^*}}
\label{nfour5}
\end{equation}
spanning ${\bf H}^n$.
Their supersymmetric partners are four spin 1/2 fermions,
whose eight components
we denote by
$\psu i  , \psut i , \psv i, \psvt i$ together with their complex conjugates
$\psus i  , \psuts i , \psvs i, \psvts i$.
On these matter fields the abelian gauge group acts
in a {\it triholomorphic} fashion, which
in our setting means  that $\{u^i,v^i\}$ have charges $\{q^i,-q^i\}$ .
The lagrangian is:
\begin{eqnarray}
\lefteqn{\cL^{(4)}_{\rm quatern}\,=
\,-\bigl(\delp u^{i^*}\,\delm u^i+\delm u^{i^*}\,\delp u^i
+\delp v^{i^*}\,\delm v^i+\delp v^{i^*}\,\delm v^i\bigr)}\nonumber\\
&&\mbox{}+4i\bigl(\psu i\,\delm\psus i +\psv i\,\delm\psvs i +\psut
i \delp\psuts i +\psvt i\,\delp\psvts i\bigr)\nonumber\\
&&\mbox{}+2i\sum_i q^i\biggl\{\biggl[\psu i\,\bigl(\lamm u^{i^*} +
\mup v^i\bigr) -\coco\biggr]-\biggl[\psv i\,\bigl(\lamm v^{i^*} -
\mup u^i\bigr) -\coco\biggr]\nonumber\\
&&\mbox{}- \biggl[\psut i\,\bigl(\latm u^{i^*} +\mutp v^i\bigr)
-\coco\biggr]+ \biggl[\psvt i\,\bigl(\latm v^{i^*} -\mutp u^i\bigr)
-\coco\biggr]\biggr\} \nonumber\\
&&\mbox{}+8i\biggl[M^*\sum_i q^i\bigl(\psu i\psuts i -\psv i\,
\psvts i\bigr)-\coco\biggr] - 8i\biggl[N\sum_i q^i\bigl(\psu i\,\psvt i
+\psv i\,\psut i\bigr)-\coco\biggr] \nonumber\\
&&\mbox{}+ 8\bigl(|M|^2
+|N|^2\bigr)\sum_i (q^i)^2\bigl(|u^i|^2 + |v^i|^2\bigr)
-2\cP\sum_i q^i\bigl(|u^i|^2 -|v^i|^2\bigr) \nonumber\\
&&\mbox{}+2i\bigl(\cQ\,\sum_i q^i
u^i v^i -\coco\bigr)
\label{nfour12}
\end{eqnarray}
Comparing with formulae (\ref{momentumcomponents}) we see that
the auxiliary field $\cP$ multiplies the
real component
${\cal D}^3(u^i,v^i)=\sum_{i} \, q^{i} \,
\left ( |u^{i}|^2 \, - \, |v^{i}|^2 \right )$,
 while $\cQ$ multiplies the holomorphic component
${\cal D}^{\ssm}(u^i,v^i) \, = -2i\, \sum_{i} \, q^{i} \, u^{i} \,
v^{i}$ of the momentum map.
\par
Consider $\cL^{(4)}=\cL^{(4)}_{\rm gauge} -\cL^{(4)}_{\rm quatern}$.
Varying in $\cP$ and $\cQ$ we obtain :
\begin{eqnarray}
\cP & = &\o 12 \biggl[r-\sum_i q^i\bigl(|u^i|^2 -|v^i|^2\bigr)\biggr] \,=
\o 12 \biggl[r-\cD^3(u,v)\biggr]\nonumber\\
\cQ & = &\o 12\biggl[s -2i\sum_i q^i u^i v^i\biggr] \,=\,\o 12
 \biggl[s-{\cal D}^{\ssp}(u,v)\biggr]
\label{nfour15}
\end{eqnarray}
Using eq.s (\ref{nfour15}) from the lagrangian we can extract the final form
of the N=4 bosonic potential. We make an exception
to our choice of focusing on U(1) gauge group
and write this potential in the case
where the gauge group is the direct sum of several U(1)'s.
This is done in order to make contact with the \hika
quotients of the next section. Use an index $a$ to enumerate
such U(1) factors; the triholomorphic
action is described by the matrices $(F^a)^i_j$
acting on the
$u^j$ and $-F^a$ on the $v^j$ fields. In practice one has to
replace $q^i\rightarrow (F^a)^i_j$
and sum over $a$. Then we get
\begin{equation}
U\, =\sum_a\biggl(\,\o 12\bigl(r_a-{\cal D}_a^3\bigr)^2
+\o 12|s_a-{\cal D}_a^{\ssp}|^2 +8\bigl(|M_a|^2
+|N_a|^2\bigr)\,\sum_{i,j} (F^a F^a)_{ij} \bigl(u^{i^*}u^j
+v^{i^*}v^j\bigr)\biggr)
\label{nfour16}
\end{equation}
As we see, the parameters $r,s$ of the N=4 Fayet-Iliopoulos
term are identified with the
levels of the triholomorphic momentum-map.

Minimizing the potential (\ref{nfour16})
we find only a N=4 $\sigma$-model phase;
the reason is the absence of an N=4 analogue of the Landau-Ginzburg potential.
Besides $M=N=0$, we must impose
$\cD^3(u,v) = r$ and $\cD^{\ssp}=s$. Taking into account
the gauge invariance of the Lagrangian, this means that the classical vacua
are characterized by having $M=N=0$ and the matter fields $u, v$ lying on
the HyperK\"ahler quotient
\begin{equation}
\label{13.1}
\cM = \cD_3^{-1}(r)\cap\cD_{\ssp}^{-1}(s) / U(1)
\end{equation}
of the quaternionic space ${\bf H}^n$ spanned by the fields $u^i,
v^i$ with respect to the triholomorphic action of the $U(1)$ gauge
group.
Considering the fluctuations around this vacuum, we can see that the fields
of the gauge multiplet are massive, together with the modes of the matter
fields not tangent to $\cM$. The low-energy theory will turn out to be the
N=4 $\sigma$-model on $\cM$.\par

It is useful at this point, in order to compare
with the N=2 case, to note that N=4 theories are nothing else
but  particular N=2 theories whose
structure allows the existence of additional supersymmetries.
\par
Thus if we look at the N=4 theory described above from an N=2 point of view,
it contains one gauge multiplet  and $2n \, +\, 1$ chiral multiplets, whose
bosonic components we denote as $\{X^A\}$, $A = 0,\ldots n$.
They are explicitely
$\{X^0=2 N,u^i,v^i\}$, with charges $\{0,q^i, -q^i\}$.
The Landau Ginzburg potential is
\begin{equation}
W\left ( \, X^{A} \, \right ) =-\,\o 14 X^0 \, \left ( \, s^* \, -
 \,\cD^{\ssm} (u,v) \, \right )
= -\,\o 14 X^0 \, \left ( \, s^* \, +2
 \, i\, \sum_i \, q^i u^{i} \, v^{i} \, \right )
\label{inducedsuperpotential}
\end{equation}
where $\cD^{\ssm} (u,v) \,=-2i\,\sum_i \, q^i u^{i} \,v^{i} \,$
is the holomorphic
part of the momentum map for the triholomorphic action of the gauge group on
${\bf H}^{n}\equiv {\bf C}^{2n}$.
The Landau-Ginzburg potential being given by
eq. (\ref{inducedsuperpotential}), the form (\ref{ntwo19}) of the N=2
bosonic potential reduces exactly to the potential of eq.
(\ref{nfour16}) (for U(1) gauge group):
\begin{eqnarray}
\label{section13bospot}
U &=& \o 12 \left(r-\cD^3\right)^2
+\sum_A |\partial_AW|^2 +8 |M|^2 \sum_i (q^i)^2
\left(|u^i|^2+|v^i|^2\right)\nonumber\\
&=& \o 12 \left(r-\cD^3\right)^2 +\o 12
|s-\cD^{\ssp}|^2 +(8|M|^2 +2 |X^0|^2)\sum_i (q^i)^2
\left(|u^i|^2+|v^i|^2\right)
\end{eqnarray}
{}From this N=2 point of view we  do not see two
different phases in the structure of the classical vacuum
because of the expression of $\cD^3(u,v)$,
see eq. (\ref{nfour15}). It is indeed clear that by
setting
\begin{equation}
\label{section13realeq}
r-\sum_i q^i\left(|u^i|^2-|v^i|^2\right)=0
\end{equation}
the exchange of $r>0$ with $r<0$ just corresponds to the exchange of the $u$'s
with the $v$'s. Everywhere else the $u$'s and the $v$'s appear
symmetrically, hence the two phases $r>0$ and $r<0$ are actually the same
thing.
This is far from being accidental. The
charge of $v^i$ is opposite to the one of $u^i$ because
of the triholomorphicity
of the action of the gauge group, which is essential in a N=4 theory; thus
the indistinguishability of the two phases is intrinsic to any N=4
theory of the type we are considering in this paper.
\par
To complete the definition of the vacuum, we must set $M=X^0=0$ and
require $\cD^{\ssp}(u,v)=s$.
\par
We now examine the reconstruction of the low-energy theory.
We again focus on the
bosonic sector.
We utilize the above N=2 point of view, and proceed as before. We just insist
now on the peculiarities due to the
form of the potential (\ref{section13bospot}).

By letting the gauge coupling constant go to infinity, we
eliminate the gauge kinetic terms and \
also that for $X^0$, since it originates from the N=4
gauge multiplet.

Note that the holomorphic contraint $\cD^{\ssp}=s$ is not implemented in the
N=2 lagrangian we are starting from through a Lagrange
multiplier. This would be the case (by means of the auxiliary field $\cQ$) had
we chosen to utilize the N=4 formalism, see eq.(\ref{nfour12}), and this is the
case for the real constraint $\cD^3=r$, through the auxiliary field $\cP$.
This fact causes no problem, as it is
perfectely consistent with what happens, from the
geometrical point of view, taking the \hika quotient. Indeed the
\hika quotient procedure is schematically represented by
\begin{equation}
\label{hikaquot}
\cS\hskip 3pt\stackrel{j^{\ssp}}{\longleftarrow}\hskip 3pt \cD_{\ssp}^{-
1}(s) \hskip 3pt\stackrel{j^3}{\longleftarrow}\hskip 3pt\cN\equiv
\cD_3^{-1}(r)\cap\cD_{\ssp}^{-1}(s)\hskip 3pt\stackrel{p}{\longrightarrow}
\hskip 3pt \cM\equiv\cN /G
\end{equation}
where we have extended in an obvious way the notation of eq. (\ref{kq1}):
$j^{\ssp}$ and $j^3$ are the inclusion maps and $p$ the projection on the
quotient.\par
We already remarked that the surface $\cD_3^{-1}(r)$ is not
invariant under the action of the {\it complexified} gauge group $G^c$.
Instead it is easy to verify that the holomorphic surface $\cD_{\ssp}^{-
1}(s)$ {\it is} invariant under the action of $G^c$. Just as in the
K\"ahler quotient procedure  we can therefore replace
the restriction to $\cD_3^{-1}(r)$ and the $G$ quotient with a $G^c$
quotient, without modifying the need of taking the restriction
to $\cD_{\ssp}^{-1}(s)$. The \hika quotient can be realized as
follows:
\begin{equation}
\label{hikaquot2}
\cS\hskip 3pt\stackrel{j^{\ssp}}{\longleftarrow}\cD_{\ssp}^{-1}(s)\hskip 3pt
\stackrel{p^c}{\longrightarrow}\hskip 3pt\cM\equiv\cD_{\ssp}^{-1}(s) /G^c
\end{equation}
We see that, in any case, we have to implement the constraint $\cD^{\ssp}=
s$.
This does not affect  the procedure of extending the
action of the gauge group to its complexification. Setting now, to be
simple and definite, $q^i=1$ $\forall i$
\footnote{This corresponds to the obvious
N=4 generalization of ${\bf CP}^N$. The spaces obtained by means
of the hyperK\"ahler quotient procedure of ${\bf H}^n$ with respect to this
$U(1)$ action have real dimension $4(n-1)$; the K\"ahler metric metric they
inherit from the quotient construction are called
Calabi metrics \cite{momentmap_3}},
the complexified group acts as:
\begin{eqnarray}
\label{section13compl}
u^i\hskip 3pt\longrightarrow\hskip 3pt e^{i\Phi}u^i \hskip 12pt &;&
\hskip 12pt v^i\hskip 3pt\longrightarrow\hskip 3pt e^{-i\Phi}v^i\nonumber\\
v\hskip 3pt &\longrightarrow&\hskip 3pt v+\o i2 (\Phi -\Phi^*)
\end{eqnarray}
One obtains the invariance of the lagrangian under this action by
means of the substitutions
$u^i\rightarrow e^{-v}u^i$, $v^i\rightarrow e^{-v}v^i$.
The variation, after these replacements,
in the auxiliary field $\cP$ gives  the equation
$\cD^3(e^{-v}u, e^v v)=r$, that is
\begin{equation}
\label{degree2eq}
r-e^{-2v}\sum_i |u^i|^2 + e^{2v}\sum_i |v^i|^2 =0
\end{equation}
This is easily solved for $v$.
We have still to implement the holomorphic constraint $\cD_{\ssp}=s$. This
task is simplified by the ${\bf C}^*$ gauge invariance of our lagrangian.
As it is clear from
the form of the ${\bf C}^*$-transformations one can for instance choose the
gauge $u^n = v^n$.
Solving explicitely the constraint gives the $\{u^i,v^i\}$, $i = 1,\ldots n$
in terms of some irreducible coordinates
$\{\hat u^I,\hat v^I\}$,$I=1,\ldots n-1$.
The final result of the appropriate manipulations that should be made on the
lagrangian will be the
reconstruction of the action for the N=2 $\sigma$-model
having as target space the \hika quotient ${\bf H}^n /U(1)$,
endowed with the K\"ahler metric which is naturally provided by this
onstruction, exactly ias it happened in the
K\"ahler quotient case. The K\"ahler quotient is again
obtained through eq. (\ref{kq4}).
It is convenient to call $\beta = \sum_i |u^i|^2$ and $\gamma = \sum_i |v^i|^2$
(both to be considered as functions of $\hat u^I,\hat v^I$).
Differently from the ${\bf CP}^N$ case, the part of the target space K\"ahler
potential  coming  from the restriction of the K\"ahler potential
for ${\bf H}^n$ to the
surface $\cD_3^{-1}(r)\cap\cD_{\ssp}^{-1}(s)$ is not an irrelevant constant.
Indeed it is given (see section  1) by:
\begin{equation}
\label{kprestricted}
K|_{\cN}=\o 12 (e^{-2v}\sum_i |u^i|^2 + e^{2v}\sum_i |v^i|^2) =\o 12
\sqrt{\rho^4 +4\beta\gamma}
\end{equation}
The final expression of the K\"ahler potential for the Calabi metric is:
\begin{equation}
\label{kpcalabi}
\hat K = \o 12\sqrt{\rho^4 +4\beta\gamma} +\o{\rho^2}2 \log \o{-\rho^2 +
\sqrt{\rho^4+4\beta\gamma}}{2\gamma}
\end{equation}
In the case $n=2$, the target space has $4$ real dimensions and the Calabi
metric is nothing else that the Eguchi-Hanson metric, i.e. the simplest
Asymptotically Locally Euclidean (ALE) gravitational instanton
\cite{momentmap_4,momentmap_2,momentmap_5}.

\section{ALE spaces as hyperK\"ahler quotients}
The most natural gravitational analogues of the Yang-Mills instantons
are geodesically complete Riemannian four-manifolds with
(anti)selfdual curvature 2-form
\footnote{In 4 dimensions (anti)selfduality
of the curvature implies that the spaces  are \hika and that
their metric automatically satisfies the vacuum Einstein equations}.
One would like
their metric  to approach  the euclidean metric at infinity,
in agreement with the ``intuitive'' picture of instantons as being
localized in finite regions of space-time. This behaviour is
however possible only modulo an additional subtlety: the base manifold has a
boundary at infinity $S^3/\Gamma$, $\Gamma$ being a
finite group of identifications. ``Outside the core of the
instanton'' the manifold looks like ${\bf
R}^4/\Gamma$ instead of ${\bf R}^4$.
This is the reason of the name given to these spaces: the
asymptotic behaviour is only {\it locally} euclidean. The unique {\it globally}
euclidean gravitational instanton is euclidean four-space itself, wich has
boundary $S^3$.
\par
The simplest of the ALE metrics is the  Eguchi-Hanson metric
\cite{momentmap_4},
which corresponds to the case where the boundary  infinity is $S^3/{\bf Z}_2$.
The so-called Multi-center metrics \cite{gravinstant_3,gravinstant_5}
correspond to the cases $S^3/{\bf Z}_n$.
The general picture \cite{gravinstant_5,momentmap_5} is as follows:
every ALE space is determined by its group of identifications $\Gamma$,
which must be a finite Kleinian subgroup of SU(2).
Kronheimer described indeed
manifolds having such a boundary; he showed that in principle a
unique selfdual metric can be obtained for each of these manifolds
\cite{momentmap_5} and, moreover, that  every selfdual metric approaching
asymptotically the euclidean one can be recovered in such a manner
\cite{hyperquotient_3}.\par
\underline{\sl The Kleinian subgroups of SU(2)}\par\noindent
Choosing complex coordinates $z_1=x-iy,z_2=t+iz$ on
${\bf R}^4\sim{\bf C}^2$, and representing a point $(z_1,z_2)$ by a
quaternion, the group
${\rm SO(4)}\sim {\rm SU(2)}_L\times {\rm SU(2)}_R$, which is
the isometry group of the sphere at infinity, acts on the
quaternion by matrix multiplication:
\begin{equation}
\twomat{z_1}{i\bar z_2}{iz_2}{\bar z_1}
\hspace{3pt}\longrightarrow\hspace{3pt} M_1 \cdot
\twomat{z_1}{i\bar z_2}{iz_2}{\bar z_1}
\cdot M_2
\label{matrixaction}
\end{equation}
the element $M\in {\rm SO(4)}$
being represented as $(M_1\in {\rm SU(2)}_L, M_2\in
{\rm SU(2)}_R)$.
The group $\Gamma$ can be seen as a finite subgroup of ${\rm SU(2)}_L$,
acting on ${\bf C}^2$ in the natural way by its two-dimensional
representation:
\begin{equation}
\label{grouptheory3}
\forall U\in\Gamma\subset {\rm SU(2)},
\hspace{15pt} U:\hspace{3pt}
{\bf v}=\twovec{z_1}{z_2}
\hspace{3pt} \longrightarrow
U{\bf v} =\twomat {\alpha}{i \beta}{i{\bar \beta}}{\bar \alpha}
\twovec {z_1} {z_2}\, .
\end{equation}
It is a classic result that the possible
finite subgroups of SU(2) are organized in two
infinite series and three exceptional cases; each subgroup $\Gamma$ is in
correspondence with a simply laced Lie algebra $\cG$, and we write
$\Gamma(\cG)$ for it. See fig.s 1,2 for the explicit correspondence.
\par
The  2-dimensional defining representation ${\cal Q}$ is  obtained
by regarding the group $\Gamma$ as an SU(2) subgroup [that is, $\cQ$
is the representation which acts in eq.(\ref{grouptheory3})].
\par
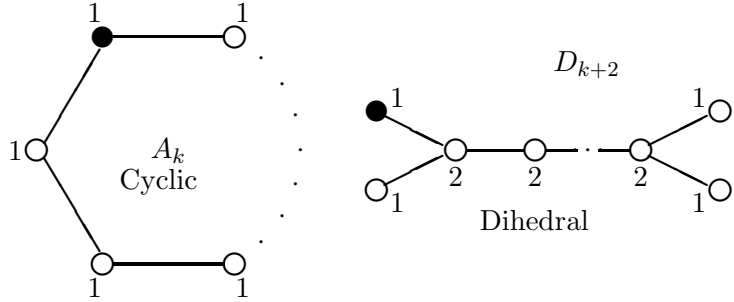
\begin{figure}[tb]
\label{dynfigure1}
\begin{center}
\begin{picture}(280,110)(-60,-60)
\thicklines
\put(25,43){\circle{8}}
\put(-25,43){\circle*{8}}
\put(-50,0){\circle{8}}
\put(-25,-43){\circle{8}}
\put(25,-43){\circle{8}}
\put(35,35){\makebox(0,0){$\cdot$}}
\put(43,25){\makebox(0,0){$\cdot$}}
\put(48,13){\makebox(0,0){$\cdot$}}
\put(50,0){\makebox(0,0){$\cdot$}}
\put(48,-13){\makebox(0,0){$\cdot$}}
\put(43,-25){\makebox(0,0){$\cdot$}}
\put(35,-35){\makebox(0,0){$\cdot$}}
\put(-21,43){\line(1,0){42}}
\put(-47,3){\line(3,5){21}}
\put(-47,-3){\line(3,-5){21}}
\put(-21,-43){\line(1,0){42}}
\put(28,52){\makebox(0,0){1}}
\put(-28,52){\makebox(0,0){1}}
\put(-58,0){\makebox(0,0){1}}
\put(-28,-52){\makebox(0,0){1}}
\put(28,-52){\makebox(0,0){1}}
\put(0,0){\makebox(0,0){$A_{k}$}}
\put(-2,-12){\makebox(0,0){Cyclic }}

\put(75,-30){
\begin{picture}(150,50)(0,-30)
\thicklines
\put(67,30){$D_{k+2}$}
\put(39,-30){Dihedral }
\multiput(30,0)(30,0){2}{\circle{8}}
\put(30,-10){\makebox(0,0){2}}
\put(60,-10){\makebox(0,0){2}}
\put(34,0){\line(1,0){22}}
\put(64,0){\line(1,0){12}}
\put(80,0){\makebox(0,0){$\cdots$}}
\put(84,0){\line(1,0){12}}
\put(100,0){\circle{8}}
\put(100,-10){\makebox(0,0){2}}
\put(0,15){\circle*{8}}
\put(8,20){\makebox(0,0){1}}
\put(0,-15){\circle{8}}
\put(8,-20){\makebox(0,0){1}}
\multiput(130,15)(0,-30){2}{\circle{8}}
\put(122,20){\makebox(0,0){1}}
\put(122,-20){\makebox(0,0){1}}
\put(3,13){\line(2,-1){22}}
\put(3,-13){\line(2,1){22}}
\put(103,2){\line(2,1){22}}
\put(103,-2){\line(2,-1){22}}
\end{picture}}
\end{picture}
\caption{\sl Extended Dynkin diagrams of the infinite series}
\end{center}
\end{figure}
\begin{figure}[tb]
\label{dynfigure2}
\begin{center}
\begin{picture}(300,135)(0,-10)
\thicklines
\put(5,25){$E_8\leftrightarrow {\cal I}$}
\put(70,18){Icosahedron}
\multiput(0,0)(30,0){7}{\circle{8}}
\put(210,0){\circle*{8}}
\put(0,-10){\makebox(0,0){2}}
\put(30,-10){\makebox(0,0){4}}
\put(60,-10){\makebox(0,0){6}}
\put(90,-10){\makebox(0,0){5}}
\put(120,-10){\makebox(0,0){4}}
\put(150,-10){\makebox(0,0){3}}
\put(180,-10){\makebox(0,0){2}}
\put(210,-10){\makebox(0,0){1}}
\multiput(4,0)(30,0){7}{\line(1,0){22}}
\put(60,4){\line(0,1){22}}
\put(60,30){\circle{8}}
\put(52,30){\makebox(0,0){3}}

\put(140,45){
\begin{picture}(180,50)(0,0)
\thicklines
\put(5,25){$E_7\leftrightarrow {\cal O}$}
\put(100,20){Octahedron}
\multiput(0,0)(30,0){6}{\circle{8}}
\put(180,0){\circle*{8}}
\put(0,-10){\makebox(0,0){1}}
\put(30,-10){\makebox(0,0){2}}
\put(60,-10){\makebox(0,0){3}}
\put(90,-10){\makebox(0,0){4}}
\put(120,-10){\makebox(0,0){3}}
\put(150,-10){\makebox(0,0){2}}
\put(180,-10){\makebox(0,0){1}}
\multiput(4,0)(30,0){6}{\line(1,0){22}}
\put(90,4){\line(0,1){22}}
\put(90,30){\circle{8}}
\put(82,30){\makebox(0,0){2}}
\end{picture}}

\put(-20,80){
\begin{picture}(120,50)(0,15)
\thicklines
\put(5,25){$E_6\leftrightarrow {\cal T}$}
\put(70,35){Tetrahedron}
\multiput(0,0)(30,0){5}{\circle{8}}
\put(0,-10){\makebox(0,0){1}}
\put(30,-10){\makebox(0,0){2}}
\put(60,-10){\makebox(0,0){3}}
\put(90,-10){\makebox(0,0){2}}
\put(120,-10){\makebox(0,0){1}}
\multiput(4,0)(30,0){4}{\line(1,0){22}}
\put(60,4){\line(0,1){22}}
\put(60,30){\circle{8}}
\put(52,30){\makebox(0,0){2}}
\put(52,60){\makebox(0,0){1}}
\put(60,34){\line(0,1){22}}
\put(60,60){\circle*{8}}
\end{picture}}
\end{picture}
\caption{\sl Exceptional extended Dynkin diagrams}
\end{center}
\end{figure}
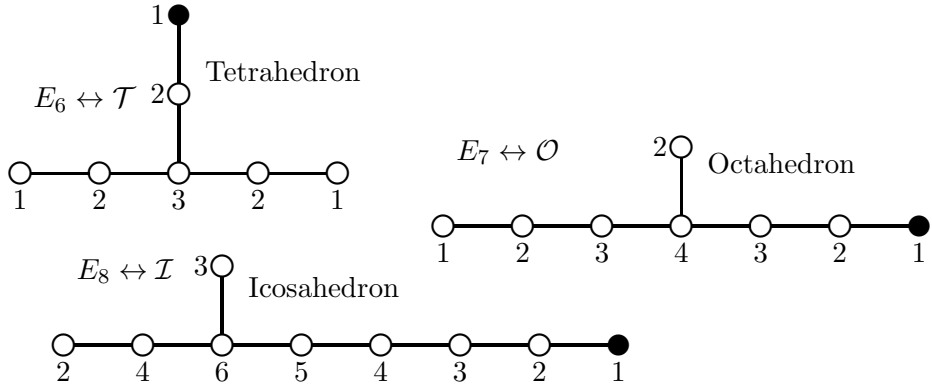

For the Kleinian groups $\Gamma$ it is
particularly important
the decomposition of the tensor product of an irreducible
representation $D_\mu$ with the
defining 2-dimensional representation ${\cal Q}$.  It is indeed at the
level of this decomposition that the relation between these groups
and the simply laced Dynkin diagrams is more explicit \cite{gravinstant_6}.
Furthermore this decomposition plays a
crucial role in the explicit construction of the ALE manifolds
\cite{momentmap_5}. Setting
\begin{equation}
{\cal Q} \, \otimes \, D_\mu ~=~\bigoplus_{\nu =0}^{r} \, A_{\mu \nu} \, D_\nu
\label{grouptheory16}
\end{equation}
where $D_0$ denotes the identity representation,
one finds that the matrix ${\bar c}_{\mu\nu}=2\delta_{\mu\nu}-A_{\mu\nu}$
is the {\it extended Cartan matrix} encoded in the {\it  extended
Dynkin diagram} corresponding
to the given group.
Recall that the extended Dynkin diagram contains in addition to the {\it  dots}
representing  the {\it  simple roots}
$\left \{ \, \alpha_1 \, ......\, \alpha_r \,  \right \}$
an {\it  extra dot} (marked black in Fig.s \ref{dynfigure1},
\ref{dynfigure2}) representing the negative of
the highest root $\alpha_0 \, = \, \sum_{i=1}^{r} \, n_i \, \alpha_i$
($n_i$ are the Coxeter
numbers). There is thus a correspondence between the non-trivial conjugacy
classes  or equivalently the non-trivial irrepses of the group
$\Gamma(\cG)$ and the simple roots of $\cG$; in this correspondence,
the extended Cartan matrix provides us with the Clebsch-Gordan coefficients
(\ref{grouptheory16}), while
the Coxeter numbers $n_i$ express the dimensions of the
irreducible representations.
All these informations are summarized in Fig.s 2,3 where the
numbers $n_i$ are attached to each of the dots; the extra dot stands for the
identity representation.

\vskip 0.2cm\noindent
$\bullet$\underline{Example}\hskip 4pt{\small\it
Consider the cyclic subgroups of SU(2), that is the $A_{k}$-series.
The defining 2-dimensional representation ${\cal Q}$ is given
by  the matrices
\begin{equation}
\gamma_l \, \in \, \Gamma (A_{k})~~~~; ~~~~\gamma_l \, = \,
{\cal Q}_l \, \eqdef \,\twomat {e^{2\pi i l/(k+1)}}{0}{0}
{e^{-\, 2\pi i l/(k+1}} \,\{ l=1,.....,k\}\, .
\label{grouptheory7}
\end{equation}
It is not irreducible since all irreducible representations
are one-dimensional as one sees from Fig. 2.
In the $j$-th irreducible representation the $l$-th element of the group is
represented by $D^{(j)}( e_l)=\nu^{jl}$, where
$\nu={\rm exp} 2\pi/k$, $j=1,\ldots k$ .
The $(k+1)\times (k+1)$ array of phases $\nu^{jl}$ appearing in
the above equation is the character table.
Given the ${\bf C}^2$ carrier space of the defining representation
[see eq.s (\ref{grouptheory3})]
one can construct three algebraic invariants, namely
\begin{equation}
z=z_1 \, z_2 \hskip 0.2cm ;\hskip 0.2cm
x=\left ( z_1 \right )^{k+1} \hskip 0.2cm ;\hskip 0.2cm
y=\left ( z_2 \right )^{k+1}
\label{grouptheory19}
\end{equation}
that satisfy the polynomial relation
\begin{equation}
W_{A_k} \, \left ( \,  x, \, y, \, z \right ) \, \eqdef \,
x \, y \, - z^{k+1} \, = \, 0 \,. \hskip 20pt\Box
\label{grouptheory20}
\end{equation}
}
\vskip 0.2cm
Analogous invariants and polynomials (see table 1) can be constructed for
the other kleinian subgroups.
\par
\underline{\sl ALE manifolds and resolution of simple singularities}
\par\noindent
The polynomial constraint
$W_{\Gamma} (x,y,z) \, = \, 0$ plays an important role in the construction
of the ALE manifolds.
Indeed, as we are going to see, the vanishing
locus in ${\bf C}^3$ of the {\it potential} $W_{\Gamma}(x,y,z)$ coincides
with the space of equivalence classes ${\bf C}^2 / \Gamma$,
that is with the singular orbifold
limit of the self-dual manifold ${\cal M}_{\Gamma}$. According to the
standard procedure of deforming singularities
\cite{singlit_1,arnold,singlit_2,singlit_3} there is a corresponding family of
smooth manifolds ${\cal M}_{\Gamma}\left(t_1, t_2 , ....., t_r \right)$
obtained as the vanishing locus  ${\bf C}^3$ of a
{\it deformed  potential}:
\begin{equation}
{\tilde W}_{\Gamma} \left (  x,y,z ; \, t_1, t_2 , .....,
t_r \right )~=~W_{\Gamma}(x,y,z)\, + \,\sum_{\alpha =1}^{r} \ t_{\alpha} \,
{\cal P}^{(\alpha}(x,y,z)
\label{grouptheory21}
\end{equation}
where $t_{\alpha}$ are complex numbers (the moduli of the complex
structure of ${\cal M}_{\Gamma}$)
and ${\cal P}^{(\alpha)}(x,y,z)$ is a basis spanning the chiral ring
\begin{equation}
{\cal R}=\o{{\bf C}[x,y,z]}{\partial W}
\label{grouptheory22}
\end{equation}
of polynomials in $x,y,z$ that do not vanish upon use of the
vanishing relations $\partial_x \, W
\,= \, \partial_y \, W \, = \, \partial_z  \, W \, = \, 0$.
The dimension of this chiral ring $| {\cal R} |$ is
precisely equal to the number of non-trivial conjugacy classes
(or of non trivial irreducible representations)  of the finite group $\Gamma$.
{}From the geometrical point of view this implies an
identification between the number of complex
structure deformations of the ALE manifold and the number $r$ of
non-trivial conjugacy classes discussed above.
\par
In this framework one can describe the homology of the
manifolds $\cM_{\Gamma}(t_{\alpha})$ with $t\not= 0$.
The non-contractible two-cycles $c_{\alpha}, \alpha=1,\ldots
r$ (each isomorphic to a copy of $\c{\bf P}^1$) can be put into correspondence
with the vertices of the non-extended Dynkin diagram for $\Gamma$.
The intersection matrix of the $c_{\alpha}$
is the negative of the Cartan matrix:
\begin{equation} c_{\alpha}\cdot c_{\beta}=\bar c_{\alpha \beta}\, .
\label{cartan}\end{equation}
The Kronheimer construction, that we shortly describe, shows that
the base manifold (simply denoted as $\cM$)
of an ALE space is diffeomorphic to the space $\cM_{\Gamma}(t_{\alpha})$
supporting the resolution of the orbifold $\csg$. Therefore
the equation (\ref{cartan}) applies to the generators of its second homology
group. In particular we see that
\begin{eqnarray}
\tau & = &dim H^2_{\bf c}(\cM)=dim H_2(\cM)=\nonumber\\
     & = &{\rm rank\,\, of\,\, the\,\, corresponding\,\, Lie\,\,
          Algebra}=\nonumber\\
     & = &\#\,\,{\rm non\,\, trivial\,\, conj.\,\, classes\,\,
          in}\,\,\Gamma= |{\cal R}|\label{tau} \, .
\end{eqnarray}
where $\tau$ is the Hirzebruch signature
\footnote{for ALE manifolds $\tau = \chi -1$, $\chi$ being the
Euler characteristic; it just counts the normalizable selfdual forms}.
These results are summarized in Table \ref{kleinale}.
\par
\begin{table}\caption{\sl KLEINIAN GROUP versus ALE MANIFOLD properties }
\label{kleinale}
\begin{center}
\begin{tabular}{||l|c|c|c|c|c||}\hline
$\Gamma$. & $W(x,y,z)$ & ${\cal R}=\o{{\bf C}[x,y,z]}{\partial W}$
&$|{\cal R}|$ & ${\#} c.~c. $& $\tau\equiv\chi -1$  \\ \hline
\hline
$A_k$&$xy - z^{k+1}$&$\{ 1, z,.. $&$k$&$k+1$&$k$ \\
$~$&$~$&  $.., z^{k-1} \}$&$~$&$~$&$~$\\ \hline
$D_{k+2}$&$x^2 +y^2 z + z^{k+1}$&$\{ 1, y, z,y^2,$&$k+2$&$k+3$&$k+2$ \\
$~$&$~$&  $ z^2, ..., z^{k-1} \}$&$~$&$~$&$~$ \\ \hline
$E_6=$&$x^2+y^3 +z^{4}$&$\{ 1, y,  z,$&$6$&$7$&$6$ \\
${\cal T}$&$~$&  $yz, z^2,yz^{2} \}$&$~$&$~$&$~$ \\ \hline
$E_7=$&$x^2+y^3 +yz^{3}$&$\{ 1, y, z,y^2,$&$7$&$8$&$7$ \\
${\cal O}$&$~$&  $z^2,yz,y^2z \}$&$~$&$~$&$~$ \\ \hline
$E_8=$&$x^2+y^3 + z^{5}$&$\{ 1,y, z,z^2,yz,$&$8$&$9$&$8$ \\
${\cal I}$&$~$&  $z^3,yz^2,yz^3  \}$&$~$&$~$&$~$ \\ \hline
\end{tabular}\end{center}
\end{table}
\par
\underline{\sl Kronheimer construction}
\par\noindent
The \hika quotient is performed on a suitable flat \hika space $\cS$
that now we define. Given any finite subgroup of SU(2), $\Gamma$, consider a
space $\cP$ whose elements are two-vectors of $|\Gamma |\times |\Gamma |$
complex matrices: $p\in \cP =\left(A, B\right)$.
The action of an element $\gamma\in \Gamma$ on the points of $\cP$ is
the following:
\begin{equation}
\twovec{A}{B} \hskip 10pt \stackrel{\gamma}{\longrightarrow}\hskip 10pt
\twomat{u_{\gamma}}{i\bar v_{\gamma}}{iv_{\gamma}}{\bar u_{\gamma}}
\twovec{R(\gamma)AR(\gamma^{-1})}{R(\gamma)BR(\gamma^{-1})}
\label{gammaactiononp}
\end{equation}
where the two dimensional matrix in the r.h.s. is the realization of
$\gamma$ in the defining representation $\cQ$ of $\Gamma$, while $R(\gamma)$
is the regular, $|\Gamma|$-dimensional
representation
\footnote{The basis vectors $e_\gamma$ of the
regular representation $R$ are in one-to-one
correspondence with the group elements $\gamma$ and transform as
$R(\gamma) e_\delta = e_{\gamma \cdot \delta}$, $ \forall
\gamma , \delta \in \Gamma$.}.
This transformation property identifies $\cP$, from the
point of view of the representations of $\Gamma$, as $\cQ\otimes {\rm End}(R)$.
The space $\cP$ can be given a quaternionic structure,
representing its elements as ``quaternions of matrices'':
\begin{equation}
p\in\cP=\twomat{A}{iB^{\dagger}}{iB}{A^{\dagger}}\hskip 1cm A,B\in {\rm
End}(R)\,\, .
\label{pquaternions}
\end{equation}
The space $\cS$ is the subspace of $\Gamma$-invariant elements in $\cP$:
\begin{equation}
\cS\eqdef\left\{p\in\cP / \forall \gamma\in\Gamma, \gamma\cdot p =
p\right\}\,\, .
\label{mdefinition}
\end{equation}
Explicitly the invariance condition  reads:
\begin{equation}
\twomat{u_{\gamma}}{i\bar v_{\gamma}}{iv_{\gamma}}{\bar u_{\gamma}}
\twovec{A}{B} \hskip 5pt =\hskip 5pt
\twovec{R(\gamma^{-1})AR(\gamma)}{R(\gamma^{-1})BR(\gamma)}\,\, .
\label{invariancecond}
\end{equation}
The space $\cS$ is elegantly described for all $\Gamma$'s using the
associated Dynkin diagram.

A two-vector of matrices can be thought of also as a matrix of
two-vectors: that is, $\cP=\cQ\otimes{\rm Hom}(R,R)={\rm
Hom}(R,\cQ\otimes R)$. Decomposing into irreducible
representations the regular
representation, $R=\bigoplus_{\nu=0}^{r} n_{\mu} D_{\mu}$,
using eq.(\ref{grouptheory16}) and the Schur's lemma, one gets
\begin{equation}
\label{defmuastratta}
\cS=\bigoplus_{\mu,\nu} A_{\mu,\nu}{\rm
Hom}(\c^{n_{\mu}},\c^{n_{\nu}})\, .
\end{equation}
The dimensions of the irrepses,  $n_{\mu}$ are expressed in
Fig.s (\ref{dynfigure1},\ref{dynfigure2}).
{}From eq.(\ref{defmuastratta}) the real dimension of
$\cS$ follows immediately: $\dim\, \cS=\sum_{\mu,\nu}2 A_{\mu\nu}
n_{\mu}n_{\nu}$ implies, recalling that $A=2\un-\bar c$ [see
eq.(\ref{grouptheory16})] and that for the  extended Cartan matrix
$\bar c n =0$,
\begin{equation}
\label{dimm}
\dim\, \cS=4\sum_{\mu}n_{\mu}^2= 4 |\Gamma |\,\, .
\end{equation}

The quaternionic structure of $\cS$ can be seen by simply writing its
elements as in eq.(\ref{pquaternions}) with $A,B$ satisfying the invariance
condition eq.(\ref{invariancecond}). Then the \hika forms and the
metric are described by $\Theta=\tr (d\bar m\wedge m)$ and
$ds^2\un=\tr(d\bar m\otimes d m)$. The
trace is taken over the matrices belonging to ${\rm End}(R)$ in each
entry of the quaternion.
\vskip 0.2cm
$\bullet$\underline{Example}\hskip 4pt{\small\it
The space $\cS$ can be easily described when $\Gamma$ is
the cyclic group $A_{k-1}$. The order of $A_{k-1}$ is $k$;
the abstract multiplication
table is that of ${\bf Z}_k$ and from it  we can immediately read off the
matrices of the regular representation.
One has $R(e_1)_{lm} = \delta_{l,m + 1}$
and of course $R(e_j)=(R(e_1))^j$.
Actually, the invariance condition eq.(\ref{invariancecond})
is best solved by changing
basis so as to diagonalize the regular representation.
Let $\nu=e^{2\pi i\over k}$, so that $\nu^k=1$. The
change of basis is performed by the matrix $S_{ij}={\nu^{-ij} \over
\sqrt{k}}$; in the new basis
$ R(e_j)={\rm diag}(1,\nu^j,\nu^{2j},\ldots,\nu^{(k-1)j})$.
The entries are the representatives of
$e_j$ in the unidimensional irrepses.

The explicit solution of eq.(\ref{invariancecond}) is
given in the above basis by
\begin{equation}
\label{invariantck}
(A)_{lm} = \delta_{l,m+1} u^l \hskip 0.3cm ; \hskip 0.3cm
(B)_{lm} = \delta_{l,m-1} v^l
\end{equation}
We see that these matrices are parametrized in terms of $2k$ complex,
i.e. $4k=4|A_{k-1}|$ real parameters.\hskip 4pt $\Box$
}
\vskip 0.2cm
Consider the action of $SU(|\Gamma |)$ on $\cP$ given, using the quaternionic
notation for the elements of $\cP$, by
\begin{equation}
\forall g\in SU(|\Gamma |),
g: \twomat{A}{iB^{\dagger}}{iB}{A^{\dagger}}
\longmapsto \twomat{gAg^{-1}}{igB^{\dagger}g^{-1}}{igBg^{-1}}
{gA^{\dagger}g^{-1}}\,\, .
\label{sunaction}
\end{equation}
It is easy to see that this action is a
triholomorphic isometry of $\cP$:\hskip 10pt $ds^2$ and $\Theta$
are invariant.
Let $F$ be the subgroup of $SU(|\Gamma |)$ which {\it commutes with the
action of $\Gamma$ on $\cP$}, the $\Gamma$-action described in
eq.(\ref{gammaactiononp}).
Then the action
of $F$ descends to $\cS\subset\cP$ to give a
{\it triholomorphic isometry}: the metric and \hika forms on $\cS$ are just
the restriction of those on $\cP$. It is therefore possible to take the
\hika quotient of $\cS$ with respect to $F$.

Let $\{f_A\}$ be a basis of generators for $\cF$, the Lie algebra of $F$. Under
the infinitesimal action of $f=\un+\lambda^A f_A\in F$,
the variation of $m\in \cS$
is $\delta m= \lambda^A\delta_A m$, with
\begin{equation}
\delta_A m = \twomat{[f_A,A]}{i[f_A,B^{\dagger}]}{i[f_A,B]}
{[f_A,A^{\dagger}]}\,\, .
\label{deltaam}
\end{equation}
The components of the momentum map (see (\ref{momentumcomponents}))
are then given by
\begin{equation}
\cD_A=\tr\,(\bar m\,\delta_A m)\,\,\,\eqdef\,\,\,
\tr\,\twomat{f_A\,\cD_3(m)}{f_A\,\cD_-(m)}{f_A\,\cD_+(m)}{f_A\,\cD_3(m)}
\label{momentummatrix}
\end{equation}
so that the real and holomorphic maps $\cD_3:\cS\rightarrow\cF^*$ and
$\cD_+:\cS\rightarrow\c\times\cF^*$ can be represented as matrix-valued maps:
\footnote{It is easy to see that indeed the matrices
$[A,A^{\dagger}]+[B,B^{\dagger}]$ and $[A,B]$ belong to the Lie algebra
of traceless matrices $\cF$; practically we identify $\cF^*$ with $\cF$
by means of the Killing metric.}
\begin{eqnarray}
\cD_3(m)&=&-i\left([A,A^{\dagger}]+[B,B^{\dagger}]\right)\nonumber\\
\cD_+(m)&=&\left([A,B]\right)\,\, .
\label{momentums}
\end{eqnarray}
Let $\cZ\equiv\cZ^*$ be the dual of the centre of $\cF$.
In correspondence with a level $\zeta=\{\zeta^3,\zeta^+\}\in{\bf
R}^3\otimes\cZ$ we can form the \hika quotient
$\cM_{\zeta}\eqdef\cD^{-1}(\zeta)/F$.
{\it Varying $\zeta$ and $\Gamma$ every
ALE space can be obtained as $\cM_{\zeta}$}.

First of all, it is not difficult to check that $\cM_{\zeta}$ is
four-dimensional. As for the space $\cS$, there is a nice characterization
of the group $F$ in terms of the extended Dynkin diagram associated
with
$\Gamma$:
\begin{equation}
F=\bigotimes_{\mu} U(n_{\mu})\,\, .
\label{formofF}
\end{equation}
One must however set
the determinant of the elements to one, since $F\subset SU(|\Gamma |)$.
$F$ has a $U(n_{\mu})$ factor for each dot of the diagram, $n_{\mu}$ being
associated to the dot as in Fig.s 1,2.
$F$ acts on the various
``components'' of $\cS$ [which are in correspondence with the edges of
the diagram, see eq.(\ref{defmuastratta})] as dictated by the structure
diagram. From
eq.(\ref{formofF}) it is immediate to derive that $\dim\, F=\sum_{\mu}
n_{\mu}^2 -1 = |\Gamma |-1$. It follows that
\begin{equation}
\dim \cM_{\zeta}=\dim\, \cS -4\, \dim F = 4|\Gamma | -4(|\Gamma
|-1)=4\,\, .
\label{dimxzeta}
\end{equation}
\vskip 0.2cm
$\bullet$\underline{Example}\hskip 4pt{\small\it
The structure of $F$ and the momentum map for its action are very simply
worked out in the $A_{k-1}$ case. An element $f$ of $F$ must commute with the
action of $A_{k-1}$ on $\cP$, eq.(\ref{gammaactiononp}),
where the two-dimensional
representation in the l.h.s. is given in eq.(\ref{grouptheory7}).
Then $f$ must have the form
\begin{equation}
f={\rm diag} (e^{i\varphi_0},e^{i\varphi_1},\ldots ,e^{i\varphi_{k-1}})
\hskip 0.2cm ; \hskip 0.2cm \sum \varphi_{i}=0\,\, .
\label{Fforck}
\end{equation}
Thus $\cF$ is just the algebra of diagonal traceless $k$-dimensional
matrices, which is $k-1$-dimensional. Choose a basis of generators for
$\cF$, for instance $f_1={\rm diag}
(1,-1,\ldots)$,$f_2={\rm diag} (1,0,-1,\ldots)$,$\ldots$,$f_{k-1}={\rm diag}
(1,0,\ldots,-1)$. From eq.(\ref{momentums}) one gets directly the components
of the momentum map:
\begin{eqnarray}
\cD_{3,A}&=&|u^0|^2-|v_0|^2-|u^{k-1}|^2-|v_{k-1}|^2
-|u^A|^2-|v_A|^2+|u^{A-1}|^2-|v_{A-1}|^2\nonumber \\
\cD_{+,A}&=&u^0 v_o - u^{k-1}v_{k-1}
- u^A v_A + u^{A-1} v_{A-1}\,\,. \hskip 20pt \Box
\label{momentummapck}
\end{eqnarray}
}
\vskip 0.2cm
In order for $\cM_{\zeta}$ to be a manifold, it is necessary that $F$ acts
freely on $\cD^{-1}(\zeta)$. This happens or not depending on the value
of $\zeta\in\cZ$.
Again, a simple characterization of $\cZ$ can be given in terms of the
simple Lie algebra $\cG$ associated with
$\Gamma$ \cite{momentmap_5}. There exists an
isomorphism between $\cZ$ and the Cartan subalgebra $\cH$ of $\cG$. Thus
we have
\begin{equation}
\dim\, \cZ=\dim\,\cH ={\rm rank}\,\cG
= \#{\rm of\,\,\,non\,\,\,trivial\,\,\,conj.\,\,\,classes\,\,\,in}\,\,\,\Gamma
\,\, .
\end{equation}
The space $\cM_{\zeta}$ turns out to be singular when, under the above
identification $\cZ\sim\cH$, any of the level components $\zeta^i\in {\bf
R}^3\otimes \cZ$ lies on the walls of a Weyl chamber.
In particular, as the point $\zeta^i=0$ for all $i$ is identified with
the origin in the root space {\it the space $\cM_0$ is singular}.
We will see in a moment that $\cM_0$ corresponds to the {\it
orbifold limit} $\csg$ of a family of ALE manifolds with boundary at
infinity $S^3/\Gamma$.

To see that this is general, choose the natural basis $\{e_{\delta}\}$ for the
regular representation $R$.
Define then the space $L\subset \cS$ as follows:
\begin{equation}
L=\left\{\twovec{C}{D}\in\cS\,/\,C,D \,\,{\rm
are\,\,diagonal\,\,in\,\,the\,\,basis\,\,}\left\{e_{\delta}\right\}
\right\}\,\, .
\label{thespacel}
\end{equation}
For every element $\gamma\in\Gamma$ there is a pair of numbers
$(c_{\gamma},d_{\gamma})$
given by the corresponding entries of $C,D$:
$C\cdot e_{\gamma}=c_{\gamma}e_{\gamma}$, $D\cdot e_{\gamma}=d_{\gamma}
e_{\gamma}$. Applying the invariance condition eq.(\ref{invariancecond}),
which is valid since $L\subset\cS$, it results that
\begin{equation}
\twovec{c_{\gamma\cdot\delta}}{d_{\gamma\cdot\delta}}=
\twomat{u_{\gamma}}{i\bar v_{\gamma}}{iv_{\gamma}}{\bar u_{\gamma}}
\twovec{c_{\delta}}{d_{\delta}}\,\, .
\label{orbitofthepair}
\end{equation}
We can identify $L$ with $\c^2$ associating for instance $(C,D)\in L
\longmapsto (c_0,d_0)\in \c^2$. Indeed all the other pairs $(c_{\gamma},
d_{\gamma})$ are determined in terms of eq.(\ref{orbitofthepair}) once
$(c_0,d_0)$ are given. By eq.(\ref{orbitofthepair}) the action
of $\Gamma$ on $L$ induces exactly the action of $\Gamma$ on $\c^2$ that
we considered in eq.s (\ref{matrixaction},\ref{grouptheory3}).
\par
Note that we can directly realize $\csg$ as an affine
algebraic surface in $\c^3$ (see eq. (\ref{grouptheory20}))
by expressing the coordinates $x$, $y$ and $z$ of $\c^3$
in terms of the matrices $(C,D)\in L$.
\vskip 0.2cm
$\bullet$\underline{Example}\hskip 4pt{\small\it
The explicit parametrization of the matrices in ${\cal S}$ in the
$A_{k-1}$case
(which was given in eq.(\ref{invariantck}) in the basis in which the regular
representation $R$ is diagonal), can be
conveniently rewritten in the ``natural'' basis $\left\{e_{\gamma}
\right\}$ via the matrix $S^{-1}$. The subset
$L$ of diagonal matrices $(C,D)$ is given by
\begin{equation}
C=c_0\, {\rm diag}(1,\nu,\nu^2,\ldots,\nu^{k-1}),\hskip 12pt
D=d_0\, {\rm diag}(1,\nu^{k-1},\nu^{k-2},\ldots,\nu),
\label{unouno}
\end{equation}
where $\nu={\rm e}^{2\pi i\over k}$. This is nothing but the fact that
$\c^2\sim L$. The set of pairs
$(\nu^m c_0,$ $\nu^{k-m}d_0)$,
$m=0,1,\ldots,k-1$ is an orbit of $\Gamma$ in $\c^2$
and determines the corresponding orbit of $\Gamma$ in $L$.
To describe $\c^2 / A_{k-1}$
we identify $(x,y,z)\in \c^3$, such that $xy=z^k$, as
\begin{equation}
x=\det\, C \hskip 12pt ;\hskip 12pt y=\det\, D,
\hskip 12pt ;\hskip 12pt z={1\over k} \tr \,CD. \hskip 20pt \Box
\label{identifyAk}
\end{equation}
}
\vskip 0.2cm
It is quite easy to show the following fundamental fact: {\it
each orbit of $F$ in $\cD^{-1}(0)$ meets $L$ in one orbit of $\Gamma$}.
Because of the above identification between $L$ and $\c^2$, this leads
to the proof that {\it $\cM_0$ is isometric to $\csg$}.
\vskip 0.2cm
$\bullet$\underline{Example}\hskip 4pt{\small\it
Let us show explicitely in the
case of the cyclic groups the  one-to-one correspondence between the orbits of
$F$ in $\cD^{-1}(0)$ and those of $\Gamma$ in $L$.
Choose the basis where $R$ is diagonal.
Then $(A,B) \in {\cal S}$ has the form of eq. (\ref{invariantck}). Now,
the relation $xy=z^k$ (eq. (\ref{grouptheory20}))
holds true also when, in eq. (\ref{identifyAk}), the pair $(C,D)\in L$
is replaced by an element $(A,B)\in \cD^{-1}(0)$.
Indeed the elements $(A,B)\in\cD^{-1}(0)$ can be described  solving
eq. (\ref{momentums}) at zero r.h.s..
It gives $u_j=|u_0|{\rm e}^{i\phi_j}$ and $v_j=|v_0|{\rm e}^{i\psi_j}$
and $\psi_j=\Phi-
\phi_j$ $\forall j$ for a certain phase $\Phi$.
Then  we immediately check that such a pair $(A,B)\in\cD^{-1}(0)$
satisfies $xy=z^k$ if $x=\det\,A$, $y=\det\,B$ and $z=(1/k) \, \tr \,AB$.
We are left  with $k+3$ parameters (the $k$ phases
$\phi_j$, $j=0,1,\ldots k-1$,
plus the absolute values $|u_0|$ and $|v_0|$ and the phase $\Phi$).
Indeed ${\rm dim}\,\cD^{-1}(0)={\rm dim}\,{\cal M}-3 \,{\rm dim}\,F=
4|\Gamma|-3(|\Gamma|-1)=|\Gamma|+3$, where $|\Gamma|={\rm dim}\,\Gamma=k$.
\par
Now we perform the quotient of $\cD^{-1}(0)$
with respect to $ F $.
Given a set of phases $f_i$ such that $\sum_{i=0}^{k-1}f_i=0\,
{\rm mod} \, 2\pi$ and given $f={\rm diag}
({\rm e}^{if_0},{\rm e}^{if_1},\ldots,{\rm e}^{if_{k-1}})\in  F $,
the orbit of $ F $ in $\cD^{-1}(0)$ passing through
$(A, B)$ is given by
$(fAf^{-1}, fBf^{-1})$. Choosing
$f_j=f_0+j\psi+\sum_{n=0}^{j-1}\phi_n$, $j=1,\ldots,k-1$, with
$\psi=-{1\over k}\sum_{n=0}^{k-1}\phi_n$, and $f_0$ determined by
the condition $\sum_{i=0}^{k-1}f_i=0\, {\rm mod} \, 2\pi$, one has
\begin{equation}
\label{op}
(fAf^{-1})_{lm}=a_0 \delta_{l,m + 1}\hskip 0.3cm ; \hskip 0.3cm
(fBf^{-1})_{lm}=b_0 \delta_{l,m - 1}
\end{equation}
where $a_0=|u_0|{\rm e}^{i\psi}$ and $b_0=|v_0|{\rm e}^{i(\Phi-\psi)}$.
Since the phases $\phi_j$ are determined modulo $2\pi$, it follows that
$\psi$ is determined modulo $2\pi\over k$. Thus we can say
$(a_0,b_0)\in {\csg}$.
This is the one-to-one correspondence between
$\cD^{-1}(0)/F$ and $\c^2/\Gamma$.\hskip 4pt $\Box$
}
\section{Resolution of ALE singularities
$W_{\Gamma}(t^{\alpha})$ and Fayet-Iliopoulos
parameters}
\vskip 0.2cm
So far we have reviewed the main points of the
Kronheimer construction. In particular we
have shown the constructive definition of the quaternionic flat space $\cS$
and of the ``gauge group'' acting on it by triholomorphic
isometries needed to retrieve an ALE space as a \hika quotient.
That is, we have described
the necessary ingredients to specify,
according to the procedure outlined in sec. 3,
an N=4 renormalizable field theory (the {\it microscopic theory})
whose low-energy effective action (the {\it macroscopic theory})
is the sigma-model on the ALE space under consideration
\footnote{Of course, to carry out explicitely
until the end computations analogous
to those for the Calabi metrics is extremely complicated; indeed the form
of the metric that would result from this quotient is in general not known,
with the exception
of the Eguchi-Hanson case.}.
\par
We do not insist on the mathematical
proofs of the main statements of Kronheimer's work
(in particular, the identification of {\it all} ALE spaces with $\cM_{\zeta})$.
We rather choose to illustrate, in the specific case of the cyclic subgroups,
an {\it explicit} relation
between the parameters $\zeta^i\in\cZ, i=1,2,3$
of the \hika construction (the levels of
the momentum map)
and the deformation parameters $t^{\alpha}$
appearing in eq. (\ref{grouptheory21}).
We divide the $\zeta$ parameters in
$r$-parameters (the real levels of the $\cD^3$ momentum map)
and $s$-parameters (the complex levels of the
$\cD^+$ momentum map) since this was the notation utilized
in sec. 3. This relation tells us explicitely
which is the ``deformed'' potential describing
an ALE space, obtained as a \hika quotient with levels $\{r,s\}$,
as an hypersurface in  ${\bf C}^3$.
We stress that the parameters $r,s$ are coupling paramenters
(the N=4 generalizations
of Fayet-Iliopoulos parameters) in the ``microscopic'' N=4 lagrangian while the
$t^{\alpha}$ are parameters in the $\sigma$-model
(the ``macroscopic'' description),
since they appear in the definition of the target space,
and in particular of its
complex structure. This gives a physical interest to the relation we describe.
\par
To find the desired relation, we have in practice to find a ``deformed''
relation between the invariants  $x,y,z$.
To this purpose, we focus on the holomorphic
part of the momentum map, i.e.\ on the equation $[A,B]=\Sigma_0$, where
$\Sigma_0={\rm diag}(s_0,s_1,\ldots,s_{k-1})$
with $s_{0}=-\sum_{i=1}^{k-1}s_i$.
Recall the expression (\ref{invariantck}) for the matrices $A$ and $B$.
Calling $a_i=u_iv_i$, $[A,B]=\Sigma_0$ implies that
$a_i=a_0+s_i$ for $i=1,\ldots,k-1$. Now,
let $\Sigma={\rm diag}(s_1,\ldots,s_{k-1})$.
We have
\begin{equation}
xy=\det A \, \det B=
a_0\, \Pi_{i=1}^{k-1}(a_0+s_i)=a_0^k\,\det \left(1+{1\over a}
\Sigma\right)=\sum_{i=0}^{k-1}a_0^{k-i}S_i(\Sigma).
\label{def1}
\end{equation}
The $S_i(\Sigma)$ are the symmetric polynomials in the eigenvalues
of $\Sigma$, defined by $\det (1+\Sigma)=
\sum_{i=0}^{k-1}S_i(\Sigma)$. In particular, $S_0=1$ and $S_1=\sum_{i=1}^{k-1}
s_i$.
Define $S_k(\Sigma)=0$, so that $xy=
\sum_{i=0}^{k}a_0^{k-i}S_i(\Sigma)$, and note that
$z={1\over k}\tr AB=a_0+{1\over k}S_1(\Sigma)$.
Then the desired deformed relation between $x$, $y$ and $z$
is obtained by substituting
$a_0=z-{1\over k}S_1(\Sigma)$ in (\ref{def1}), obtaining finally
\begin{eqnarray}
&&xy=\sum_{m=0}^k\sum_{n=0}^{k-m}\left(\matrix{k-m\cr n}\right)
\left(-{1\over k}S_1(\Sigma)\right)^{k-m-n}S_m(\Sigma) z^n=\sum_{n=0}^k
t_n z^n.\\
&&\Longrightarrow\hskip 10pt t_n=\sum_{m=0}^{k-n}\left(\matrix{k-n\cr m}\right)
\left(-{1\over k}S_1(\Sigma)\right)^{k-m-n} S(\Sigma)_n .
\label{def2}
\end{eqnarray}
Notice in particular that $t_k=1$ and $t_{k-1}=0$, i.e.\
$xy=z^k+\sum_{n=0}^{k-2}t_nz^n$, which
means that the deformation proportional to $z^{k-1}$ is absent.
This establishes a clear correspondence between the momentum
map construction and the polynomial ring ${\bf C}[x,y,z]/ \partial W$
where $W(x,y,z)=xy-z^k$ [compare with eq. ({\ref{grouptheory21})].
Moreover, note that we have only used one
of the momentum map equations, namely $[A,B]=\Sigma_0$.
The equation $[A,A^\dagger]+[B,B^\dagger]=R$ has been completely
ignored. This means that the deformation of the complex structure
is described by the parameters $\Sigma$, while the parameters
$R$ describe the deformation of the K\"ahler class.

The relation (\ref{def2}) can also be written in a simple factorized form,
namely
\begin{equation}
xy=\Pi_{i=0}^{k-1}(z-\mu_i),
\end{equation}
where
\begin{eqnarray}
\mu_i&=&{1\over k}(s_1+s_2+\cdots +s_{i-1}
-2s_i+s_{i+1}+\cdots+s_k), \,\,\,\,
i=1,\ldots,k-1\nonumber\\
\mu_0&=&-\sum_{i=1}^k\mu_i={1\over k}S_1(\Sigma).
\end{eqnarray}


\begin{thebibliography}{10}

\bibitem{topf4d_1}
D.~Anselmi and P.~Fre'.
\newblock {`` Topological sigma models in four dimensions and tri--holo morphic
  maps "}.
\newblock {\em Nucl. Phys.}, B416:255, (1994).

\bibitem{topf4d_2}
E.~Witten and N.~Seiberg.
\newblock {``Electric--magnetic duality, monopole condensation and confinement
  in N=2 supersymmetric Yang--Mills theory"}.
\newblock {\em Nucl. Phys.}, B426:19, (1994).

\bibitem{topf4d_3}
E.~Witten and N.~Seiberg.
\newblock {``Monopoles, duality and chiral supersymmetry breaking in N=2 QCD"}.
\newblock {\em hep--th}, /9408013:~~, (1994).

\bibitem{topf4d_4}
E.~Witten.
\newblock {``~Monopoles and four manifolds"}.
\newblock {\em IASSNS-HEP-9496 hep-th}, /9411102:~~, (1994).

\bibitem{topf4d_5}
A.~Ceresole, R.~{D'Auria}, and S.~Ferrara.
\newblock {``On the geometry of moduli space of vacua in N=2 supersymmetric
  Yang--Mills theory"}.
\newblock {\em CERN-TH 7384/94 POLFIS-TH}, 07/94:~~, (1994).

\bibitem{gravinstant_1}
M.~Billo', P.~Fre', L.~Girardello, and A.~Zaffaroni.
\newblock {`` Gravitational instantons in heterotic string theory: the {H}--map
  and the moduli deformations of (4,4) superconformal theories "}.
\newblock {\em Int. Jour. Mod. Phys.}, A8:2351, (1993).

\bibitem{gravinstant_2}
D.~Anselmi, M.~Billo', P.~Fre', L.~Girardello, and A.~Zaffaroni.
\newblock {`` {ALE} manifolds and conformal field theoris "}.
\newblock {\em Int. Jour. Mod. Phys.}, A9:3007, (1994).

\bibitem{topphases_1}
M.~Billo' and P.~Fre'.
\newblock {`` N=4 versus N=2 phases, {HyperK\"ahler} quotients and the 2d
  topological twist "}.
\newblock {\em Class. and Quantum Grav.}, 11:785, (1994).

\bibitem{momentmap_5}
P.B. Kronheimer.
\newblock {`` The construction of ALE spaces as HyperK\"ahler quotients "}.
\newblock {\em Jour Diff. Geo.}, 29:665, (1989).

\bibitem{gravinstant_3}
S.~Hawking and C.N. Pope.
\newblock {`` Symmetry breaking by instantons in supergravity "}.
\newblock {\em Nucl. Phys.}, B146:381, (1978).

\bibitem{miscellaneous_1}
P.~Fayet and J.~Iliopulos.
\newblock {`` Spontaneously broken supergauge symmetries and Goldstone spinors
  "}.
\newblock {\em Phys. Lett.}, 51B:46, (1974).

\bibitem{topftwist_1}
D.~Anselmi and P.~Fre'.
\newblock {`` Twisted N=2 supergravity as topological gravity in four
  dimensions "}.
\newblock {\em Nucl. Phys.}, B392:401, (1993).

\bibitem{topftwist_2}
D.~Anselmi and P.~Fre'.
\newblock {`` Topological twist in four dimensions, {R} duality and
  hyperinstantons "}.
\newblock {\em Nucl. Phys.}, B404:288, (1993).

\bibitem{momentmap_2}
N.J. Hitchin, A.~Karlhede, {U. Lindstr\"om}, and M.~{Rocek}.
\newblock {``HyperK\"ahler metrics and supersymmetry"}.
\newblock {\em Comm. Math. Phys.}, 108:535, (1987).

\bibitem{hyperquotient_2}
U.~Lindstrom and M.~Rocek.
\newblock {`` Scalar-tensor dualities and N=1,2 non-linear sigma-models "}.
\newblock {\em Nucl. Phys.}, B222:285, (1983).

\bibitem{hyperquotient_3}
P.B. Kronheimer.
\newblock {`` A Torelli-type theorem for gravitational instantons "}.
\newblock {\em Jour Diff. Geo.}, 29:685, (1989).

\bibitem{momentmap_4}
T.~Eguchi and A.J. Hanson.
\newblock {``Self--dual solutions to {Euclidean} gravity"}.
\newblock {\em Ann. Phys.}, 120:82, (1979).

\bibitem{gravinstant_4}
G.W. Gibbons and S.~Hawking.
\newblock {`` Classification of gravitational instanton symmetries "}.
\newblock {\em Comm. Math. Phys.}, 66:381, (1979).

\bibitem{gravinstant_5}
N.J.Hitchin.
\newblock {`` Polygons and gravitons "}.
\newblock {\em Math. Proc. Cambridge Phylos. Soc.}, 85:465, (1979).

\bibitem{topphases_2}
E.~Witten.
\newblock {`` Phases of N=2 theories in two dimensions "}.
\newblock {\em Nucl. Phys.}, B403:159, (1993).

\bibitem{momentmap_1}
K.~Galicki.
\newblock {``A generalization of the momentum mapping construction for
  quaternionic {K\'haler} manifolds"}.
\newblock {\em Comm. Math. Phys.}, 108:117, (1987).

\bibitem{momentmap_3}
E.~Calabi.
\newblock {``Metriques K\"ahleriennes et fibres holomorphes"}.
\newblock {\em Ann. Scie. Ec. Norm. Sup.}, 12:269, (1979).

\bibitem{gravinstant_6}
J.~McKay.
\newblock {`` Graphs, singularities and finite groups "}.
\newblock {\em Proc. Symp. Pure Math., Am. Math. Soc.}, 37:183, (1980).

\bibitem{singlit_1}
M.~Artin.
\newblock {\em Am. J. Math.}, 88:129, (1966).

\bibitem{arnold}
V.I. Arnold, S.M. Gusein-Zade, and A.N. Varchenko.
\newblock {\em Singularities of differntiable maps}.
\newblock Birkh\"auser, 1975.

\bibitem{singlit_2}
E.~Brieskorn.
\newblock in {\em Actes Congr\'es Intern. Math. Ann. (t. 2)}, (1970).

\bibitem{singlit_3}
P.~Slodowy.
\newblock {\em Simple Singularities and Simple Algebraic Groups}.
\newblock Lect. Notes in Math. 815, Springer Verlag, 1980.

\end{thebibliography}

\end{document}